**Near-Infrared Light Emitting Metal Halides: Materials, Mechanisms, and Applications**

*Ying Liu, Francesco Di Stasio, Chenghao Bi, Jibin Zhang, Zhiguo Xia*, Zhifeng Shi* and Liberato Manna**

Y. Liu, J. Zhang, Z. Shi
Key Laboratory of Materials Physics of Ministry of Education
School of Physics and Microelectronics
Zhengzhou University
Zhengzhou 450052, China
E-mail: shizf@zzu.edu.cn

F. Di Stasio
Photonic Nanomaterials
Istituto Italiano di Tecnologia
Via Morego 30, Genova 16163, Italy.

L. Manna
Nanochemistry
Istituto Italiano di Tecnologia
Via Morego 30, Genova 16163, Italy
E-mail: liberato.manna@iit.it

C. Bi
Qingdao Innovation and Development Base
Harbin Engineering University
Sansha Str. 1777, Qingdao 266500, China

Z. Xia
School of Physics and Optoelectronics
South China University of Technology
Guangzhou, 510641 China
E-mail: xiazg@scut.edu.cn

**Abstract.** Near-Infrared (NIR) light emitting metal halides are emerging as a new generation of optical materials owing to their appealing features, which include low-cost synthesis, solution processability and adjustable optical properties. NIR emitting perovskite-based light-emitting diodes (LEDs) have reached an external quantum efficiency (EQE) over 20% and a device stability of over 10,000 h. Such results have sparked an interest in exploring new NIR metal halide emitters. In this review, we summarize several different types of NIR-emitting metal halides, including lead/tin bromide/iodide perovskites, lanthanide ions doped/based metal halides, double perovskites, low dimensional hybrid and $Bi^{3+}$/$Sb^{3+}$/$Cr^{3+}$ doped metal halides, and assess their recent advancements. The characteristics and mechanisms of narrow-band or broadband NIR luminescence in all these materials are discussed in detail. We also highlight the various applications of NIR-emitting metal halides and provide an outlook for the field.

## 1. Introduction

The Near-Infrared (NIR) refers to a specific region of the electromagnetic spectrum with a wavelength range of approximately 700–1800 nm, corresponding to a frequency range of ~167–429 THz. NIR light-sources (e.g., light-emitting diodes or lasers) are of interest for a variety of applications, such as hyperspectral, three-dimensional (3D) and adverse weather imaging, night vision for surveillance and automotive safety,[1] as well as biomedical imaging and point-of-care testing.[2] Nowadays, NIR light sources are mainly built from expensive epitaxial III–V semiconductors,[3] which are difficult to integrate with complementary metal-oxide-semiconductor (CMOS) silicon electronics and suffer from low volume manufacturing.[4] Compared to III–V semiconductors, various metal halides, among them halide perovskites, possess appealing features for NIR optoelectronics, above all broader spectral tunability, lower costs of synthesis and easier CMOS integration.[5] Especially their low temperature solution processing should contribute to an easier integration with CMOS technology that can only withstand low processing temperatures (< 180 ºC).[6] In recent years, different types of NIR emitting metal halides have been reported.

Lead-free metal halides include structures with different dimensionalities.[7] Examples are three-dimensional (3D) double perovskites, two-dimensional (2D) $Cs_3Bi_2Br_9$, one-dimensional (1D) $CsMnBr_3$, and zero-dimensional (0D) $Cs_2ZrCl_6$.[8] Many of these structures are not NIR emitting *per se* but can become so upon proper doping. As a note of caution, the dimensionality here refers to the connectivity of the polyhedra in the structural framework and not to the dimensionality in morphology for the nanoscale size crystals (quantum dots—0D, nanowires—1D, nanosheets—2D). The optoelectronic properties of 3D metal halides, like for example those with the perovskite structure, given the high connectivity of the polyhedra, are essentially dictated by extent of the coupling between neighboring polyhedra in the three spatial dimensions and can be described with the same models used for the more conventional semiconductors. Also, as for the more conventional semiconductors, in these materials quantum confinement effects are observed when the size of the crystals shrinks to a few nanometers. However, given the strong polarity of the lattice in metal halides, the exciton Bohr radius is usually only a few nm large, so relevant confinement effects are seen only for ultrasmall sizes. On the other hand, the properties of low-dimensional metal halides are essentially controlled by isolated structural units (single polyhedra, layers, or wires of corners-, edge- or face-sharing polyhedra) separated from each other by large inorganic (usually $Cs^+$) or organic cations.[9]

**Figure 1** provides a classification of the various NIR emitting metal halides according to their emission band ranges (Panel a) and photophysical mechanisms (Panel b). This is given here as a quick guide for the reader, while additional details on the various mechanisms are discussed in the later sections. In the emission from the band-edge recombination, the PL originates from the recombination of band-edge electrons and holes (free excitons) that can occur in direct-bandgap halide perovskites. It is characterized by a narrow emission bandwidth and small Stokes shift. Another viable mechanism of emission is the one from *f*–*f* transitions of $Ln^{3+}$ ions. $Ln^{3+}$ ions can have several 4*f* excitation levels and thus can exhibit multiple emission lines. The 4*f* electrons occupy orbitals that are very contracted, and so they behave as core-like states, hence transitions involving electrons in these states are characterized by sharp emission lines. Quantum cutting is an optical process whereby a high-energy photon is converted into two low-energy photons, resulting in a high PLQY that can exceed 100%.[10] For example, when $Yb^{3+}$ is doped into a semiconductor with wide bandgap (> twice the value of the $Yb^{3+}$ *f*–*f* transition energy), a quantum cutting process can occur. A self-trapped exciton (STE) instead can be seen as an electron-hole pair that is localized by a field generated by a local, transient distortion of the lattice that is possible in the presence of strong electron–phonon coupling. Relaxation from an STE to the ground state occurs via interaction with various lattice vibrational modes (phonons), which explains the broad FWHM of the emitted light and the large Stokes shift.[11] Finally, there is the possibility of emission from states localized on selected ion dopants (for example $Cr^{3+}$, $Bi^+$, $Sb^{3+}$, and $Bi^{3+}$).

In this review we will first provide a bird's eye view of the field from an historical standpoint. We will then delve into the progress made in the synthesis and understanding of the various materials, which we classify based on the different photophysical processes of light emission. We will then move on to discuss their applications (mainly in LEDs and biomedical imaging). Finally, we provide an outlook on the challenges and opportunities lying ahead, especially for what concerns further improvements in light emission and device performance.

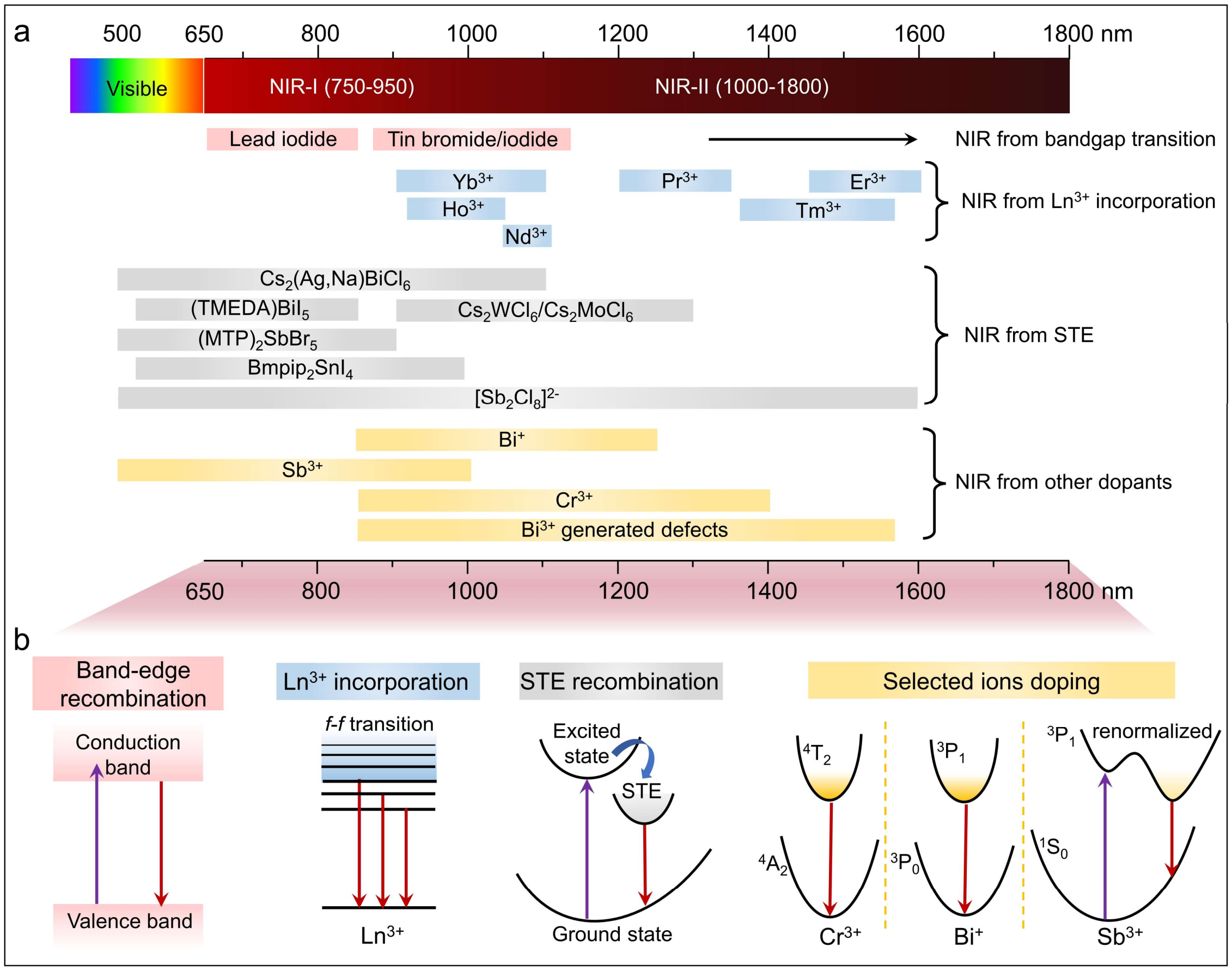

***Figure 1.*** *(a) Different possible origins of NIR emission in metal halides, including band-edge recombination, emission from $Ln^{3+}$ states, STE recombination, and emission from other types of dopants. The widths of each line indicate the range of their emission bands. For $Ln^{3+}$ ions, the figure only shows the emission band with the strongest intensity among the multiple transitions of each ion. (b) Sketches of the corresponding mechanisms of NIR emission.*

## 2. A short history of NIR emitting metal halides

NIR emitting metal halide materials have been developed mainly in the past five years, but their history goes back to decades ago. Cubic $CsSnX_3$ (where X = Cl, Br, or I) crystals have been known since 1970, when S. R. A. Bird et al. produced a black $CsSnBr_3$ cubic compound.[12] It might also be possible that such crystals were reported previously by other authors, in particular if one considers that ammonia salts of $SnBr_4I_2^{2-}$ have been first reported at the end of the 19th century.[13] These seminal works focused on studying the formation of trihalogenostannate crystals and their respective structures, while one of the first observations of photoluminescence dates back to 1995, when S. J. Clark et al. measured the emission at 720 nm from $CsSnBr_3$ crystals.[14] Interest toward NIR emitting metal halides reemerged in the early 21$^{th}$ century when organic-inorganic hybrid lead iodide perovskites with NIR emission at 700–800 nm and tin iodide perovskites with NIR emission at 800–1000 nm were reported.[15] Their compositions was $APbI_3$, in which A = $Cs^+$, methylammonium ($CH_3NH_3^+$, MA) or formamidinium ($CH(NH_2)_2^+$, FA). Thanks to their solution processability and high luminescence efficiency (photoluminescence quantum yield, PLQY > 70%), lead-based perovskites, either as films or nanocrystals (NCs), were soon embedded in NIR light-emitting devices (LEDs) with perovskites as the emitter layer.[16] The toxicity of lead and the intrinsic instability of lead-based perovskites were immediately recognized as critical roadblocks for technological exploitation. Sn has the same electronic configuration ($ns^2$) as Pb but is much less toxic. $CsSnI_3$, for example, is a long-known compound, but its optical properties were scrutinized only later in more detail. $CsSnI_3$ thin films were reported by Shum et al. in 2010, with the direct bandgap of 1.3 eV and NIR PL emission at ~950 nm.[15b] Kanatzidis et al. in 2012 synthesized $CsSnI_3$ polymorphic crystals and reported that the black orthorhombic perovskite phase had strong room-temperature NIR emission peaking at 950 nm;[15c] in 2013, the same group reported hybrid tin iodide perovskites $ASnX_3$ (A = $Cs^+$, MA or FA; X = $Br^-$ or $I^-$) single crystals with NIR emission (800–1000 nm).[15a] In 2016, the Böhm's group used the hot-injection method to prepare $CsSnX_3$ (X = $Br_{0.5}I_{0.5}$ and I) NCs with NIR emission centered at 800 and 950 nm and a maximum PLQY of 0.14% ($CsSnBr_3$ NCs).[17] Researchers have also tried to use the tin-based $MASnI_3$ perovskite to fabricate NIR LED devices, but their EQE was low (less than 1%).[18] One major problem for Sn-based perovskite is unfortunately the tendency of $Sn^{2+}$ to be oxidized to $Sn^{4+}$ under ambient conditions.[19]

To extend the NIR emission to the range of the telecommunication window (1.35–1.55 μm),[20] lanthanide ions ($Ln^{3+}$) emitting in specific spectral regions (Figure 1a) were then introduced in lead-based perovskites. The interest in $Ln^{3+}$ doped perovskites was sparked in 2017 when $Yb^{3+}$ doped $CsPbCl_{1.5}Br_{1.5}$ NCs were synthesized and were found to exhibit quantum cutting, thus achieving a high NIR PLQY (>100%).[21] Soon after, other $Ln^{3+}$ ions ($Pr^{3+}$, $Nd^{3+}$, $Ho^{3+}$ and $Er^{3+}$) were introduced as dopants in lead-based perovskites, to further extend the NIR emission range to the NIR-II region, albeit with a relatively low PLQY (1%–15%).[22] The incorporation of $Nd^{3+}$, $Pr^{3+}$ and $Er^{3+}$ in $CsPbCl_3$ films or NCs enabled intense light emissions at 1060 nm, 1300 nm, and 1540 nm, respectively. Starting from 2018, with the emergence of lead-free metal halides, $Ln^{3+}$ ions doped lead-free metal halides were also studied.[23] $Ln^{3+}$ doping ($Yb^{3+}$, $Nd^{3+}$, $Tm^{3+}$, and $Er^{3+}$) was found to yield NIR PLQYs values that could be as high as ~80%, depending on different hosts, dopants, and sample

dimensionality.[8d, 24] In these different $Ln^{3+}$ doped cases, the spectral positions of NIR luminescence peaks arising from 4*f*–4*f* forbidden transitions of one type of $Ln^{3+}$ ions are almost independent on the type of hosts and/or co-dopants ($Sb^{3+}$, $Te^{4+}$, $Bi^{3+}$, etc), due to the atomic-like transitions involving the *f* orbitals.[25] On the other hand, the absorption/excitation spectra of $Ln^{3+}$ doped perovskites are strongly dependent on the type of host (for example its bandgap), the types of defects or even the specific synthesis method, which can even influence the PLQYs of the $Ln^{3+}$ NIR emissions to a significant extent.[24a, 24c, 26] As an example, it has been reported that, during the synthesis of Yb-doped $Cs_2AgBiBr_6$ films by sequential vapor deposition, the PLQY could be optimized by controlling the substrate temperature and the $BiBr_3$ evaporation rate.[27] In most cases, $Ln^{3+}$ ions are doped into hosts with direct and relatively wide bandgap (such as lead chlorides, indium chlorides and zirconium chlorides), while narrow-bandgap perovskites appear to be more challenging hosts.[25] Also, the presence of co-dopants can increase the PLQY and lower the excitation energy of $Ln^{3+}$ NIR PL.[24a, 24c, 26] In the case of perovskite nanocrystals, a high density of surface traps can lead to a decreased PLQYs of the NIR emission compared to the corresponding $Ln^{3+}$ doped bulk materials (single crystals, powders or films).[27-28]

In certain fields, such as food analysis and medical diagnosis, broadband NIR emission (FWHM > 100 nm) is more appealing than narrow-band emission, because several functional groups (O-H, C-H, N-H) in food and human tissues are expected to interact with light at specific wavelength ranges within the NIR region. Thus, a broadband NIR light source can provide more effective and reliable analytical results.[29] Starting from 2019, when $Bmpip_2SnI_4$ (Bmpip: 1-butyl-1-methylpiperidinium) with broadband NIR emission ranging from 550 to 990 nm was synthesized,[30] several other studies have reported metal halides with broad PL band at 700–1070 nm and PLQY of 1%–75% arising from a STEs. Broadband NIR emission in metal halides can also be achieved by doping with selected ions (for example $Cr^{3+}$, $Bi^{+}$, $Bi^{3+}$, $Sb^{3+}$). The history of $Cr^{3+}$ doped metal halides with NIR emission dates back to the 1980s, when the broadband NIR luminescence at low temperature in $Cr^{3+}$ doped $Cs_2NaInCl_6$, $Cs_2NaYCl_6$, and $Cs_2NaYBr_6$ was reported.[31] To date, $Cr^{3+}$ doping induced NIR emission has been indeed obtained only in double perovskite hosts, with the PL peak at 958–1010 nm, FWHM of 165-195 nm and PLQY of 6%–23%.[32] It has been ascertained that the PL peak can be tuned by changing the crystal field, due to the strong dependence of the $Cr^{3+}$ emission on it. In a weak crystal field, $Cr^{3+}$ ions exhibit the $^4T_2 \rightarrow {}^4A_2$ spin-allowed transition with low energy. Due to the strong electron–phonon coupling of the $^4T_2$ level, the spectra will broaden.[33] The first studies on $Bi^{+}$ doping induced NIR emission in metal halides date back to 2014, with the PL peaking at 905–1015 nm and a FWHM of 117–157 nm.[34] In this case, the metal halide hosts have generally a $ABCl_3$ stoichiometry, with A = $K^{+}$, $Rb^{+}$, $Cs^{+}$, Cs; B = $Mg^{2+}$, $Cd^{2+}$.[34-35] The $Bi^{+}$ ion has $6p^2$ electronic configuration and its emissions originates from the $^3P_1 \rightarrow {}^3P_0$ transition.[36] The NIR emission from $Bi^{3+}$ dopants has been reported mainly in lead-based perovskites ($CsPbI_3$, $MAPbBr_3$ and $MAPbI_3$) since 2016, with a PL peak around 1100 nm and FWHM >170 nm.[37] Differently from the $Bi^{+}$ ion, the NIR emission from $Bi^{3+}$ dopants is ascribed to doping induced defects and the exact origins need further investigation. Broadband NIR luminescence from $Sb^{3+}$-doped metal halides has been reported since 2021.[38] This is generally characterized by a PL peak at 734–810 nm, broad FWHMs of ~170 nm and PLQYs values (8%–70%) higher than those of $Cr^{3+}$, $Bi^{+}$ and $Bi^{3+}$ doping induced NIR.[38-39] The origins of $Sb^{3+}$ doping induced NIR emission in different hosts include $^3P_1 \rightarrow {}^1S_0$ transition and possibly an STE. The detailed explanation for the NIR emission mechanisms of $Cr^{3+}$, $Bi^{+}$, $Bi^{3+}$ and $Sb^{3+}$ ions doping in different metal halides will be discussed in Section 5.2. This concludes a short historical perspective of the field, with the main milestones. What follows is a more detailed discussion on the various materials based on the mechanism of emission.

## 3. Metal halides with NIR emission from band-edge recombination

### 3.1 Lead iodide perovskites

Lead iodide perovskites were initially studied mainly having in mind applications in solar cells due to their low cost, excellent light absorption, ideal band gap, large carrier mobility and carrier diffusion length.[15a, 40] Since Friend et al. reported the surprisingly high photoluminescence quantum yields (PLQY, 70%) of $MAPbI_{3-x}Cl_x$ perovskite films in 2014, NIR-emitting lead iodide perovskites have sprung up in the field of light-emitting diodes (LEDs).[16a, 41] Lead iodide perovskites, with generic formula $APbI_3$ (A = $Cs^{+}$, $MA^{+}$ or $FA^{+}$), are characterized by small bandgaps ($CsPbI_3$: ~1.73 eV; $MAPbI_3$: ~1.63 eV; $FAPbI_3$: ~1.48 eV) and have near-infrared photoluminescence (PL) through band-edge recombination. Depending on temperature, $CsPbI_3$ is stable in various phases, including NIR-emitting perovskite phases ($\alpha$, $\beta$, $\gamma$) and a non-luminescent one ($\delta$) (**Figure 2**a). When cooled to room temperature (RT), the cubic phase ($\alpha$, $Pm\bar{3}m$) of $CsPbI_3$ will convert to the tetragonal phase ($\beta$, $P4/mbm$, 593 K), then to the 3D-orthorhombic phase ($\gamma$, $Pbnm$, 425 K) and finally to the 1D-orthorhombic non-perovskite phase ($\delta$, $Pnma$) (Figure 2a).[42] Similarly, $FAPbI_3$ at RT is more stable in the 1D hexagonal phase ($P6_3/mmc$), with no emission (Figure 2b), while $MAPbI_3$ at RT is more stable in the 3D tetragonal structure, $I4/mcm$.[43]

A simplified sketch of the electronic structure of lead iodide perovskites is shown in Figure 2c. The valence band maximum (VBM) is determined by antibonding Pb 6*s*–I 5*p* interactions, while the conduction band minimum (CBM) is mainly contributed by Pb 6*p* orbitals.[44] 1D $\delta$-$CsPbI_3$ has an indirect bandgap of 2.82 eV, larger than other 3D $CsPbI_3$ phases with direct bandgaps (1.68–1.8 eV). That is due to a higher-symmetry structure yielding stronger antibonding coupling between Pb 6*s* and I 5*p*, thus leading to a larger spread in the bandwidth of valence band and decreased bandgap.[45] The bandgaps of $APbI_3$ perovskites are also influenced by the type of A-site cations. Although these do not participate to the VBM or CBM, they affect indirectly the bandgap by regulating the tilting of the octahedra and the perovskite lattice constant.[46] A larger ionic radius of the A-site cation results in the expansion of the lattice and a narrower bandgap. The ionic radii of $Cs^+$, $MA^+$ and $FA^+$ are 1.88, 2.17 and 2.53 Å, respectively, so $FAPbI_3$ has the narrowest bandgap.[46-47] Alloying $Cs^+$ and $FA^+$ at the A site can thus be used to fine-tune the PL emission of perovskites, in addition to the broader tuning by alloying at the halide ion site. Kovalenko et al. partially replaced $FA^+$ with $Cs^+$ and $I^-$ with $Br^-$ in $FAPbI_3$ NCs (prepared by a hot-injection synthesis) thus delivering $Cs_xFA_{1-x}Pb(Br_{1-y}I_y)_3$ NCs whose PL emission peaks could be fine-tuned between 690~775 nm, while maintaining the size and shape of the pristine $FAPbI_3$ NCs.[48] Polavarapu et al. reported fast A-site cation cross-exchange at room temperature in $APbX_3$ NCs (A = $Cs^+$, $MA^+$, $FA^+$), by which they could prepare double- and triple- A-cation perovskite NCs, enabling fine tuning of the PL emission in the 680–775 nm range.[49]

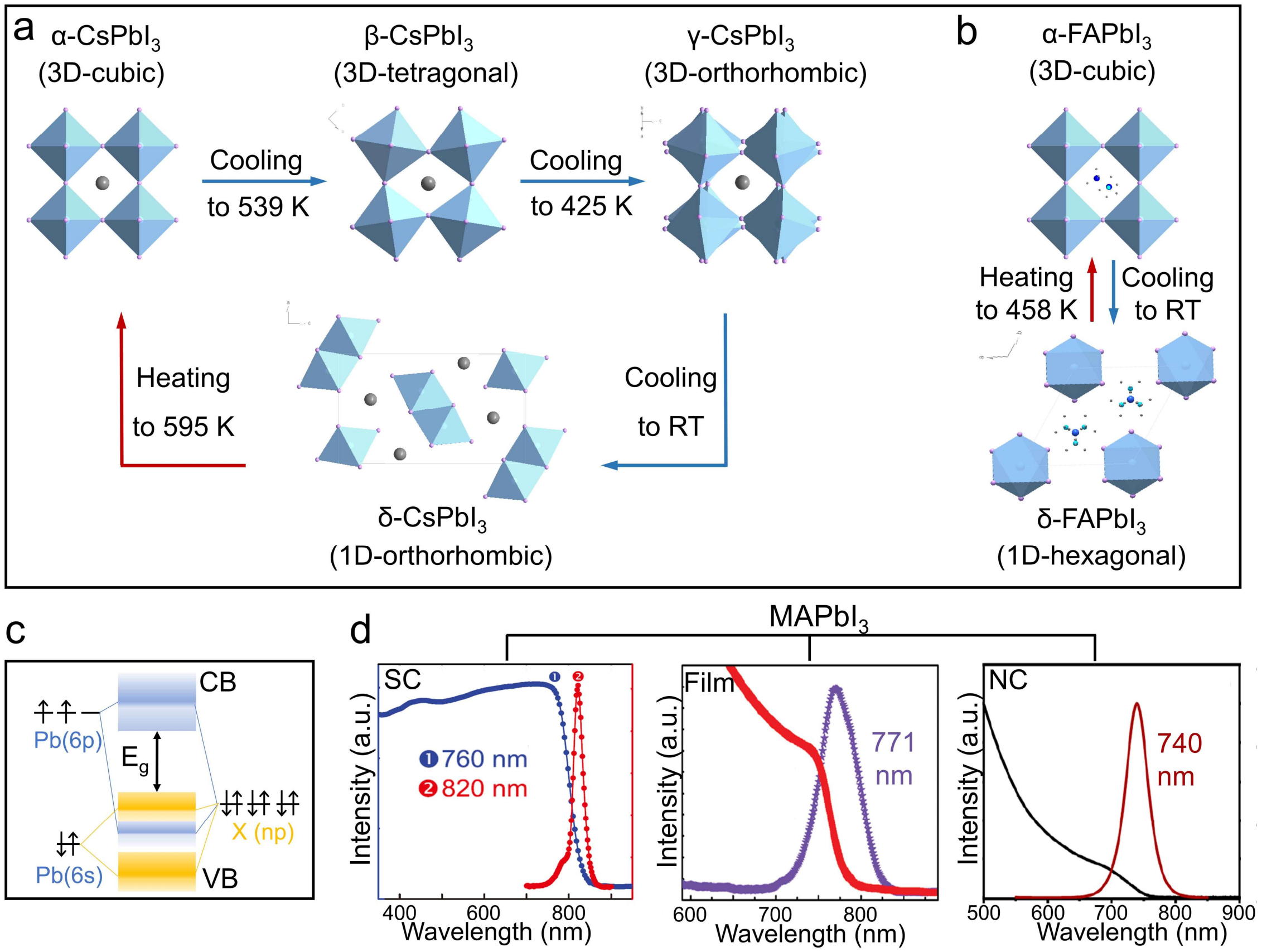

**Figure 2.** (a) Phase transitions of $\alpha$-, $\beta$-, $\gamma$-, and $\delta$-$CsPbI_3$ at different temperatures. Reproduced (Adapted) with permission.[42c] Copyright 2018, American Chemical Society. (b) Phase transition of $\alpha$- and $\delta$-$FAPbI_3$ at different temperatures. Reproduced (Adapted) with permission.[43c] Copyright 2016, Wiley. (c) Sketch of bonding/antibonding orbitals and band structure of $APbI_3$ perovskite. Reproduced (Adapted) with permission.[44] Copyright 2018, Springer Nature. (d) The absorption and PL spectra of $MAPbI_3$ single crystal, thin film, and nanocrystals. Reproduced (Adapted) with permission.[16b, 50] Left to right: Copyright 2015, AAAS; Copyright 2016, Royal Society of Chemistry; Copyright 2021, American Chemical Society.

The NIR PL properties of $APbI_3$ perovskites depend on sample dimensionality. For example, the PL emission peaks of $MAPbI_3$ single crystal, thin film and nanocrystal are at 820, 771 and 740 nm, respectively (Figure 2d).[16b, 50] The PL red-shift from thin films to single crystals stems from longer carrier diffusion length and PL reabsorption in bulk crystals.[51] Due to the small Stokes shift in $APbI_3$ perovskites, the high-energy part of the PL emitted from the inner region of single crystals is reabsorbed. Hence, in single crystals the low-energy part dominates the PL emission.[51b] Instead, in NCs their smaller sizes give rise to the quantum confinement and dielectric confinement effects, which are responsible for the blue-shifted PL.[52]

### 3.2 Tin bromide/iodide perovskites

Tin ($Sn^{2+}$) is also a typical B-site cation in the $ABI_3$ perovskite, possessing the same $ns^2$ electronic configuration as the $Pb^{2+}$ ion. Tin-based metal halides have smaller bandgaps compared to their lead-based analogues, leading to emission at longer wavelength. For instance, $CsSnI_3$ exhibits NIR emission peaking at ~950 nm with no visible region components,[15c] while $CsPbI_3$ has emission centered at ~700 nm with a tail in the visible region.[53] Similar to Pb based perovskites, Sn based perovskites possess peculiar electronic band structures, with the valence *s* orbital of Sn contributing to the valence band maximum (VBM), and the respective *p* orbitals contributing to the conduction band minimum (CBM),[44] leading to the formation of a direct bandgap. Besides, strong spin-orbit coupling (SOC) leads to an increased band dispersion and thus a suitable bandgap for NIR applications.[54] All-inorganic tin-based perovskites $CsSn(Br_{0.5}I_{0.5})_3$ and $CsSnI_3$ NCs have NIR emissions peaking at 800 and 950 nm, respectively.[17] The emission from organic-inorganic Sn analogues ($MASnI_3$ and $FASnI_3$) is in the range of 850–1000 nm.[55]

The structure and emission properties of Sn-based iodide perovskites were reported to be temperature-dependent.[15c, 55] $CsSnI_3$ undergoes various phase transitions at high temperature. Starting from RT, with increasing temperature orthorhombic $CsSnI_3$ ($\gamma$, *Pbnm*) will transform to a tetragonal phase with $P4/mbm$ ($\beta$) symmetry at 360 K and then to a cubic phase with $Pm\bar{3}m$ symmetry ($\alpha$) above 440 K (**Figure 3**a).[55] Along with the structural evolution, the PL peak of $CsSnI_3$ blue-shifts upon increasing temperature, indicating bandgap widening (Figure 3b).[55] Generally, the bandgap would decrease with increasing temperature following the thermal broadening of the electronic bands, as in the case of III–V and II–VI semiconductors. However, such thermal broadening effect is overshadowed by the displacement of $Sn^{2+}$ ions from the octahedron center upon heating. Like the band structure of lead iodide perovskites, the VBM of $CsSnI_3$ is characterized by an antibonding hybrid state of the Sn-5*s* and I-5*p* orbitals, and the CBM is mainly determined by unoccupied Sn-5*p* orbitals. Strong overlap between Sn-5*s* and I-5*p* states pushes the antibonding Sn-5*s* states up in energy, closer to the unoccupied Sn-5*p* states in the CBM. The displacement of the $Sn^{2+}$ ions upon heating reduces orbital overlap in the VBM and thus results in bandgap widening.[55-56]

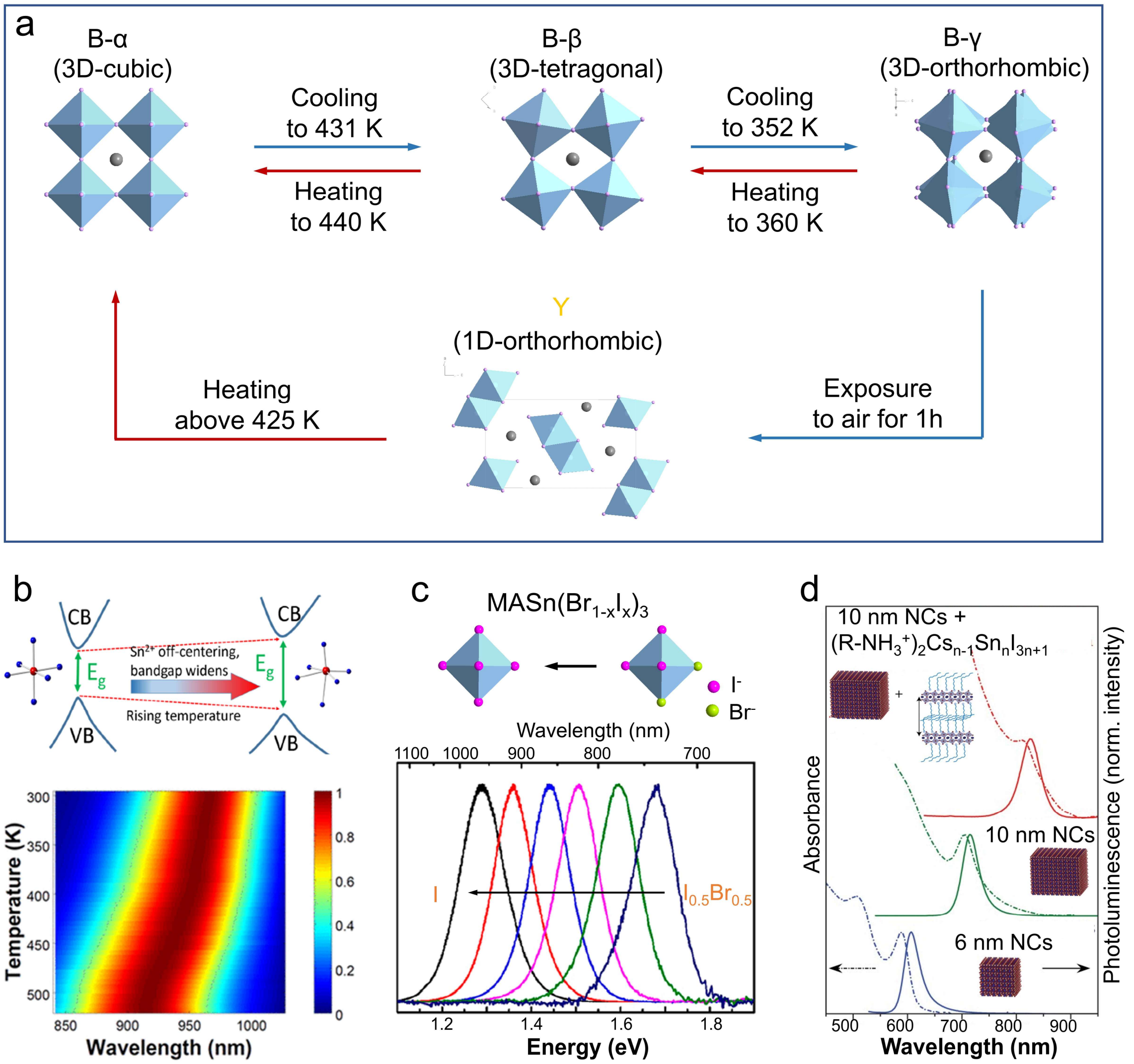

**Figure 3.** (a) Phase transitions of $\alpha$-, $\beta$-, $\gamma$-, and Y-$CsSnI_3$ at different temperatures. (b) Schematic of temperature-dependent bandgap, $Sn^{2+}$ ion position and PL spectra of $CsSnI_3$. Reproduced (Adapted) with permission.[55] Copyright 2018, American Chemical Society. (c) Normalized PL spectra $MASn(Br_{1-x}I_x)_3$ films with different halide compositions. Reproduced (Adapted) with permission.[18] Copyright 2016, American Chemical Society. (d) Optical absorptions and PL spectra of 6 nm (bottom), 10 nm $CsSnI_3$ NCs (middle) and the mixture containing 10 nm $CsSnI_3$ NCs and $(R\text{-}NH_3^+)_2Cs_{n-1}Sn_nI_{3n+1}$ nanosheets (top). Reproduced (Adapted) under the terms of the CC-BY Creative Commons Attribution 4.0 International License.[57] Copyright 2022, The Authors, Published by Wiley.

Apart from the temperature, the NIR PL emission properties of tin halide perovskites are also dependent on halide compositions and particle size. For example, as the iodide content increases, the PL emission peak of $MASn(Br_{1-x}I_x)_3$ films exhibits a red-shift (Figure 3c) and the PLQY increases from 0.1% ($MASn(Br_{0.5}I_{0.5})_3$) to 5.3% ($MASnI_3$). On the other hand, the size-dependent PL emission in NCs is due to modification of the quantum confinement upon variation of the particle size. For example, Protesescu et al. controlled the precursors and ligands ratios and prepared size-tunable $CsSnI_3$ NCs (Figure 3d).[57] The synthesized $CsSnI_3$ NCs with lateral sizes of 6 nm and 10 nm showed PL emission peaking at 606 nm and 714 nm, respectively, with a PLQY of 1%–5%. They also reported that in the process of NC nucleation and growth, before adding the Cs precursor, the byproduct 2D $(R\text{-}NH_3^+)_2SnI_4$ could form. When Cs was in excess, the formation of the 2D perovskite structures $(R\text{-}NH_3^+)_2Cs_{n-1}Sn_nI_{3n+1}$ (R = $C_{18}H_{35}$) with n > 1 was reported. By controlling the Cs:Sn and Sn:ligands ratios, a mixture of $CsSnI_3$ NCs (10 nm) and 2D $(R\text{-}NH_3^+)_2Cs_{n-1}Sn_nI_{3n+1}$ nanosheets was obtained, showing only one broad NIR PL emission centered at 825 nm (Figure 3d).[57]

As previously stated, tin iodide perovskites suffer from poor stability due to the facile oxidation of $Sn^{2+}$ ions. Many strategies have been followed to address this limitation, such as ions alloying, Sn source purification and addition of an antioxidant (typical additives are $SnF_2$ and hydroxybenzene sulfonic acid).[58] For example, Shen et al. reported that doping $Na^+$ into Sn-Pb alloyed perovskite NCs could effectively improve the PLQY from ~0.3% to 28%.[58d] That was attributed to the stronger chemical interaction between $I^-$ and $Sn^{2+}$ ions after $Na^+$ doping, so $Sn^{2+}$ ions were stabilized in the structure and the formation of $I^-$ vacancy defects was suppressed.[58d]

## 4. Metal halides with narrow-band NIR emission arising from $Ln^{3+}$ ions

### 4.1 $Ln^{3+}$ ions doped lead-based perovskites

The incorporation of $Ln^{3+}$ ions introduces new energy levels and thus endows metal halides perovskites with PL covering visible and NIR regions. As $Ln^{3+}$ ions prefer to have a relatively high coordination number (CN ≥ 6), lead-based perovskites with the octahedral cation site (CN = 6) are proper hosts for $Ln^{3+}$ ions doping. Among the $Ln^{3+}$ ions, $Yb^{3+}$ has become a popular dopant thanks to the quantum cutting mechanism when such ion is introduced in a semiconductor with sufficiently wide bandgap.[10, 22b] $Yb^{3+}$ ions possess an electronic structure including a $^2F_{5/2}$ excited state and a $^2F_{7/2}$ ground state separated by an energy difference of ~1.3 eV, corresponding to a NIR emission peaking at ~1000 nm ($^2F_{5/2} \rightarrow {}^2F_{7/2}$ transition). The energy gap of $CsPbCl_3$ matches well with twice the value of the $Yb^{3+}$ *f–f* transition energy, which can indeed lead to a quantum cutting process and thus a high NIR PLQY (>100%) for $Yb^{3+}$ doped $CsPbCl_3$. Both Song's and Gamelin's groups reported quantum cutting in $Yb^{3+}$ doped $CsPbCl_3$ NCs.[10, 21] $Yb^{3+}$ doped $CsPbCl_3$ NCs had dual emission, in the visible and NIR regions,

respectively.[10b] With more $Yb^{3+}$ dopants, the PLQY of the $CsPbCl_3$ host decreased from 20% to <1%, while the PLQY of NIR luminescence from $Yb^{3+}$ ions increased from 10% to 130% (**Figure 4**a), thus indicating energy transfer and quantum cutting processes from the host to $Yb^{3+}$.[10b] $Yb^{3+}$ was also doped into the 2D layered hybrid perovskite $(PEA)_2PbBr_4$ microcrystals (MCs), introducing a NIR emission at ~997 nm (in addition to the visible emission at 414 nm from the host), again originating from the energy transfer to the $Yb^{3+}$ ions (Figure 4b).[59] The undoped $(PEA)_2PbBr_4$ microcrystals had two PLE bands peaked at 340 nm and 407 nm and one narrow-band emission at 414 nm.[59] However, by increasing the $Yb^{3+}$ concentration, the NIR emission of $(PEA)_2PbBr_4$:$Yb^{3+}$ initially increased and then decreased, evidencing no quantum cutting effect.[59]

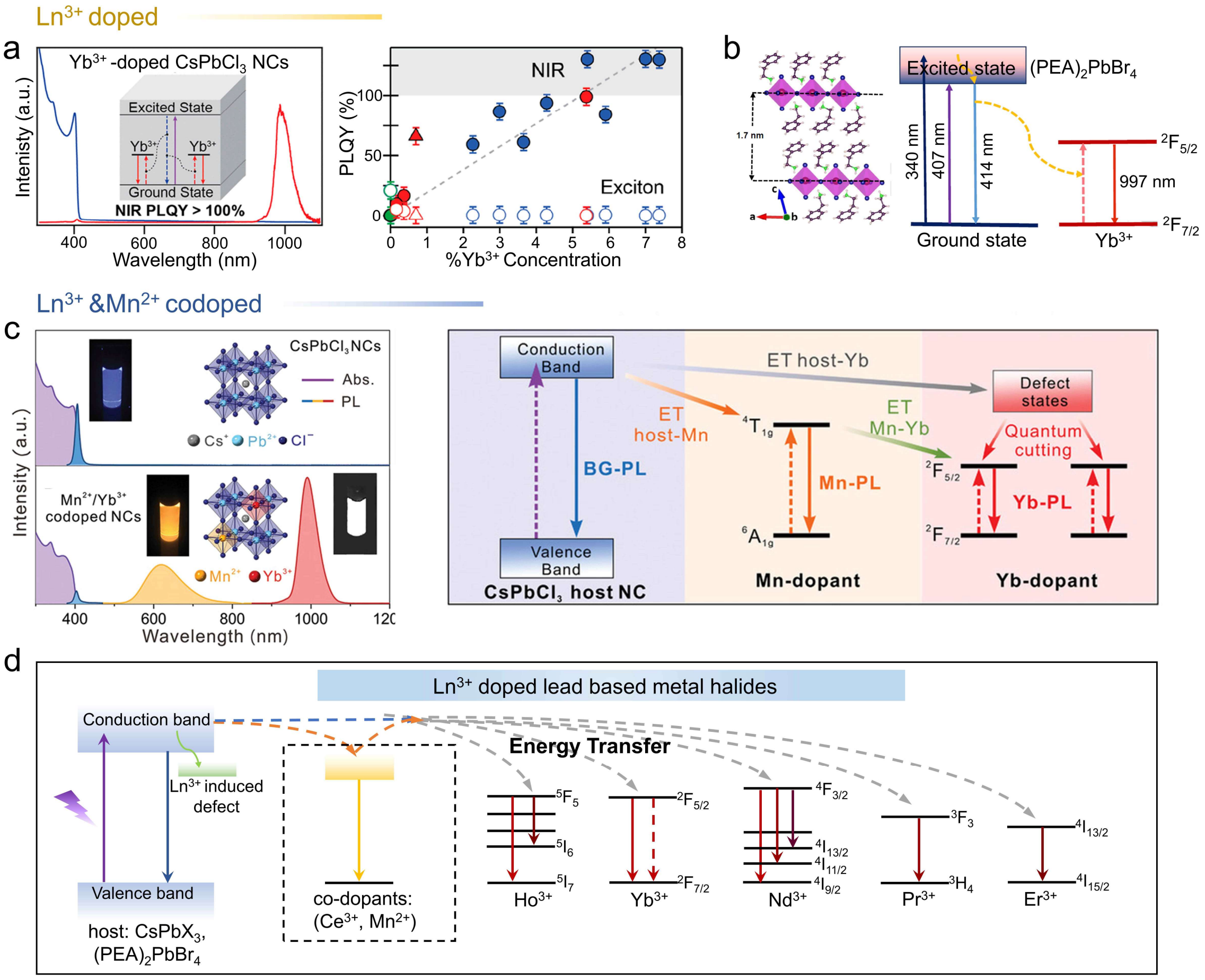

**Figure 4. $Yb^{3+}$ doped $CsPbCl_3$ NCs and $(PEA)_2PbBr_4$:** (a) Absorption and PL spectra, scheme of the quantum cutting process, and PLQYs for $CsPbCl_3$:$Yb^{3+}$ NCs. Reproduced (Adapted) with permission.[10b] Copyright 2018, American Chemical Society. (b) Structural model and scheme of emission mechanism for $Yb^{3+}$ doped $(PEA)_2PbBr_4$ microcrystals. Reproduced (Adapted) with permission.[59] Copyright 2022, Wiley. **$Yb^{3+}$/$Mn^{2+}$ co-doped $CsPbCl_3$ NCs:** (c) Structure, absorption, PL spectra and proposed models of energy migration.

Reproduced (Adapted) under the terms of the CC-BY Creative Commons Attribution 4.0 International License.[60] Copyright 2020, The Authors, published by Wiley. (d) **Summary sketches** of the energy transfer process for $Ln^{3+}$ doped or $Ln^{3+}$/other ions co-doped metal halides.

Ions co-doping can lead to multiple emissions and enhance the luminescence by virtue of energy transfer.[21-22, 60] Co-dopants can act as a bridge between the host and $Ln^{3+}$ ions with NIR emission, promoting a more efficient energy transfer and thus increase the PLQY of the NIR emission. For example, Song et al. co-doped $Ce^{3+}$ and $Yb^{3+}$ ions into $CsPbCl_{1.5}Br_{1.5}$ NCs.[21] $CsPbCl_{1.5}Br_{1.5}$:$Ce^{3+}$,$Yb^{3+}$ NCs showed three emission peaks at 455, 488 and 988 nm under the excitation of 365 nm, originating from the host, $Ce^{3+}$ ion and $Yb^{3+}$ ion, respectively. The NIR PLQY was 94% and 119% for solely $Yb^{3+}$ doped and $Ce^{3+}$/$Yb^{3+}$ co-doped NCs, respectively, thus indicating a more efficient down-conversion process in the $Ce^{3+}$/$Yb^{3+}$ co-doped samples.[21] Miyasaka et al. co-doped $Yb^{3+}$ and $Er^{3+}$ ions into a $CsPbCl_3$ film and observed $Yb^{3+}$ luminescence at 983 nm and $Er^{3+}$ luminescence at 1540 nm.[22b] The PLQY of $Er^{3+}$ emission reached up to 12.6% and was attributed to quantum cutting (with the PLQY of $Yb^{3+}$ emission above 100%) and an energy transfer process from $Yb^{3+}$ to $Er^{3+}$ ions with an efficiency over 80%.[22b] $Mn^{2+}$ ions could also act as a bridge between the host and NIR-emitting $Ln^{3+}$ ions.[22c, 60] For example, Artizzu et al. co-doped $Mn^{2+}$ and $Ln^{3+}$ into $CsPbCl_3$ NCs to activate NIR emissions.[22c] The $Mn^{2+}$ and $Ln^{3+}$ ($Nd^{3+}$, $Ho^{3+}$ and $Er^{3+}$) co-doped $CsPbCl_3$ NCs exhibited intense NIR bands at ~1060, ~1000 and ∼1500 nm, through an efficient $Mn^{2+}$-mediated energy-transfer process with a PLQY of ∼0.8%, ∼0.7% and ~3%, respectively.[22c] Chen et al. synthesized $CsPbCl_3$ NCs co-doped with $Mn^{2+}$ and $Yb^{3+}$ ions characterized by a triple-wavelength emission in the ultraviolet/blue, visible, and near-infrared regions with the highest total PLQY of ~125% (Figure 4c).[60] In addition, the PL lifetime of the emission from $Mn^{2+}$ decreased by reducing the Mn concentration, indicating a possible energy transfer process from $Mn^{2+}$ to $Yb^{3+}$ ions (Figure 4c).[60]

Co-dopants can also enhance the luminescence of the host, thus leading to an improved luminescence of NIR-emitting $Ln^{3+}$ ions through energy transfer from the host to $Ln^{3+}$.[22a] For example, Song et al. co-doped $Ni^{2+}$ and $Pr^{3+}$ ions into $CsPbCl_3$ NCs.[22a] $Ni^{2+}$ ions could passivate the defects of NCs and increase the PLQY of $CsPbCl_3$ NCs. The synthesized $CsPbCl_3$:$Pr^{3+}$, $Ni^{2+}$ NCs exhibited emission bands at 400 nm and 1300 nm with an overall PLQY of 89% and the highest infrared PLQY of 23%.[22a]

In summary, for NIR-emitting $Ln^{3+}$ doped lead-based metal halides ($CsPbX_3$ and $PEA_2PbBr_4$), different characteristic NIR luminescence bands of $Ln^{3+}$ ions ($Ho^{3+}$, $Yb^{3+}$, $Nd^{3+}$, $Pr^{3+}$ and $Er^{3+}$) can be observed by tuning the energy transfer process host → $Ln^{3+}$ or host → co-dopants → $Ln^{3+}$ (Figure 4d). Quantum cutting can take place in $Yb^{3+}$ doped $CsPbCl_3$ and the incorporation of suitable co-dopants can lead to an increased PLQY of NIR luminescence from $Ln^{3+}$. It should be noted that the host $CsPbX_3$ NCs were reported to suffer from instability under ambient atmosphere and humidity.[61] To overcome the problem, Li et al. fabricated $SiO_2$-encapsulated $CsPbBr_3$:$Yb^{3+}$ NCs in situ by hydrolyzing organosilicon (3-Triethoxysilylpropylamine, APTES).[62] The obtained $CsPbBr_3$:$Yb^{3+}$@$SiO_2$ NCs exhibited NIR emission at 985 nm with PLQY of 64% and high stability under ambient environment for over 15 days.[62]

### 4.2 $Ln^{3+}$ ions doped double perovskites

$Ln^{3+}$ ions have also been doped into more stable lead-free hosts compared with lead-based metal halides, such as double perovskites, to achieve NIR emission. Double perovskites, with the general formula $A_2B^+B^{3+}X_6$, have a 3D perovskite structure with [$B^+X_6$] and [$B^{3+}X_6$] corner-sharing octahedra. $Ln^{3+}$ ions such as $Yb^{3+}$, $Nd^{3+}$, $Er^{3+}$ and $Tm^{3+}$ ions were doped into $Cs_2AgBiX_6$ (X = $Cl^-$, $Br^-$), $Cs_2AgInCl_6$ and $Cs_2NaInCl_6$, exhibiting their characteristic NIR emissions in the 800-1500 nm spectral region.[23b, 27, 63] For example, Aydil et al. synthesized (by physical vapor deposition) $Yb^{3+}$ doped $Cs_2AgBiBr_6$ thin films having strong NIR luminescence centered at 1000 nm, with a NIR PLQY of 95%;[27] Lee et al. and Nag et al. both synthesized $Yb^{3+}$ doped $Cs_2AgInCl_6$ NCs using the modified hot-injecting method and obtained samples with NIR emission at 996 nm;[23b, 63a] Chen et al. reported $Yb^{3+}$ doped $Cs_2NaInCl_6$ with 39% NIR PLQY (**Figure 5**a).[63d] In some cases, with $Ln^{3+}$ doping, in addition to the *f–f* transitions, the $Ln^{3+}$ ion exhibits another absorption band, namely, the charge transfer band (CTB) or the ligand-to-metal charge transfer band. The CTB band is derived from the electron transfer from adjacent ligand ions to the $Ln^{3+}$ ion.[64] In $Cs_2NaInCl_6$, $Yb^{3+}$ doping introduces a new absorption band peaking at 273 nm, conforming to the charge transfer band (CTB) absorption stemming from the charge transfer from $Cl^-$ to $Yb^{3+}$. Due to the ionic character of the Na-Cl bond, weak coupling of Na 3s orbitals and Cl 3p orbitals promotes the $Cl^-$–$Yb^{3+}$ charge transfer process in $Cs_2NaInCl_6$:$Yb^{3+}$, resulting in the localization of electrons on $[YbCl_6]^{3-}$ octahedra and thus high NIR PLQY (Figure 5a).[63d]

**Table 1.** PL performance of $Ln^{3+}$ doped double perovskites.

| Host | Morphology | Doped Ln | Host $\lambda_{em}$ (nm) | NIR $\lambda_{em}$ (nm) | NIR $\lambda_{ex}$ (nm) | NIR PLQY (%) | Ref. |
|---|---|---|---|---|---|---|---|
| $Cs_2AgInCl_6$ | NCs | Yb/Er | 395 | 996/1537 | 300 | 3.6/0.05 | [23b] |
| | MCs/NCs | Yb | /~607 | 994 | 300/270 | / | [63a] |
| $Cs_2AgInCl_6$:Bi | MCs | Yb/Er | ~620 | 994/1540 | 370 | / | [24a] |
| | MCs | Nb/Er/Tm/Yb | 610 | 901,1077,1367/814,998,1543/807,1227,1477/991 | 350 | / | [63c] |
| $Cs_2AgIn_{0.99}Bi_{0.01}Cl_6$ | NCs | Pr/Ho/Tm/Er/Yb | ~630 | 1046/890,1074,1357/806,1222/808,1540/986 | 365 | 56.7(Nd) | [65] |
| $Cs_2AgInCl_6$:Sb | MCs | Nd/Yb | 650 | 1000/1000 | 365 | / | [63b] |
| $Cs_2AgBiX_6$ (X = $Cl^-$, $Br^-$) | NCs | Yb | 680 | 1000 | 365 | / | [63e] |
| $Cs_2AgBiBr_6$ | film | Yb | ~620 | 1000 | 430 | / | [66] |
| | film | Yb | / | 997 | 360 | 95 | [27] |
| | film | Yb | 630 | 997 | 420 | 82.5 | [67] |
| $Cs_2NaInCl_6$ | SCs | Yb/Yb&Er | 450 | 994/1540 | 273 | 39.4/7.9 | [63d] |
| $Cs_2Na_{0.6}Ag_{0.4}InCl_6$:Bi | MCs | Yb | 560 | 996 | 340 | / | [68] |
| $Cs_2Na_{0.2}Ag_{0.8}InCl_6$ | SCs | Yb | 640 | 996 | 365 | 46 | [69] |
| $Cs_2Na_{0.8}Ag_{0.2}BiCl_6$ | MCs | Yb/Er | 680 | 995/1540 | 360 | 19/4.3 | [70] |
| $Cs_2Na_{0.4}Ag_{0.6}InCl_6$:0.3%Bi | SCs | Er | 610 | 1540 | 365 | 38.3 | [71] |

Despite the promising results associated to NIR-emitting $Ln^{3+}$ doped double perovskites, due to low absorption coefficient and parity-forbidden *f–f* transitions, their NIR emission has limited efficiency.[23a] Table 1 presents the PL properties of NIR emitting $Ln^{3+}$ doped double perovskites. Except for $Yb^{3+}$ doped $Cs_2AgBiBr_6$ films with 95% PLQY, the highest NIR PLQY values of $Yb^{3+}$ doped $Cs_2AgInCl_6$ and $Cs_2NaInCl_6$ were 3.6% and 39.4%, respectively, far below the PLQY of $Ln^{3+}$ ions doped lead-based perovskites.[23b, 63a, 63b] To enhance the NIR luminescence, n$s^2$ ions, including $Bi^{3+}$, $Sb^{3+}$ and $Te^{4+}$ ions, were co-doped with $Ln^{3+}$ ions in double perovskite $Cs_2AgInCl_6$ or $Cs_2NaInCl_6$.[24a, 26, 63b, 63c, 65, 72] The incorporation of $Bi^{3+}$, $Sb^{3+}$ or $Te^{4+}$ can give rise to new absorption channels. The increased absorbed energy can be effectively transferred to the $Ln^{3+}$ ions, thus promoting the NIR emission from $Ln^{3+}$ ions. For example, Nag et al. co-doped $Bi^{3+}$ into $Cs_2AgInCl_6$:$Yb^{3+}$ and $Cs_2AgInCl_6$:$Er^{3+}$ microcrystals to enhance the NIR emission.[24a] $Bi^{3+}$ doping/co-doping could elicit emission from STE in $Cs_2AgInCl_6$ (Figure 5b). Besides, $Bi^{3+}$–$Er^{3+}$ co-doping provided a new absorption band at longer wavelength (350–400 nm) and increased the NIR PL intensity 45-fold at 1540 nm compared with $Er^{3+}$ doped $Cs_2AgInCl_6$ (Figure 5b).[24a] Similarly, the same group reported $Sb^{3+}$–$Er^{3+}$-co-doped and $Te^{4+}$–$Ln^{3+}$ ($Er^{3+}$, $Yb^{3+}$)-co-doped $Cs_2NaInCl_6$ microcrystals. Here, the $5s^2 \rightarrow 5s^15p^1$ transitions of $Sb^{3+}$ and $Te^{4+}$ endow additional absorptions at around 322 nm and 400 nm, respectively, thus enhancing the NIR emission of $Ln^{3+}$ ions.[26, 72]

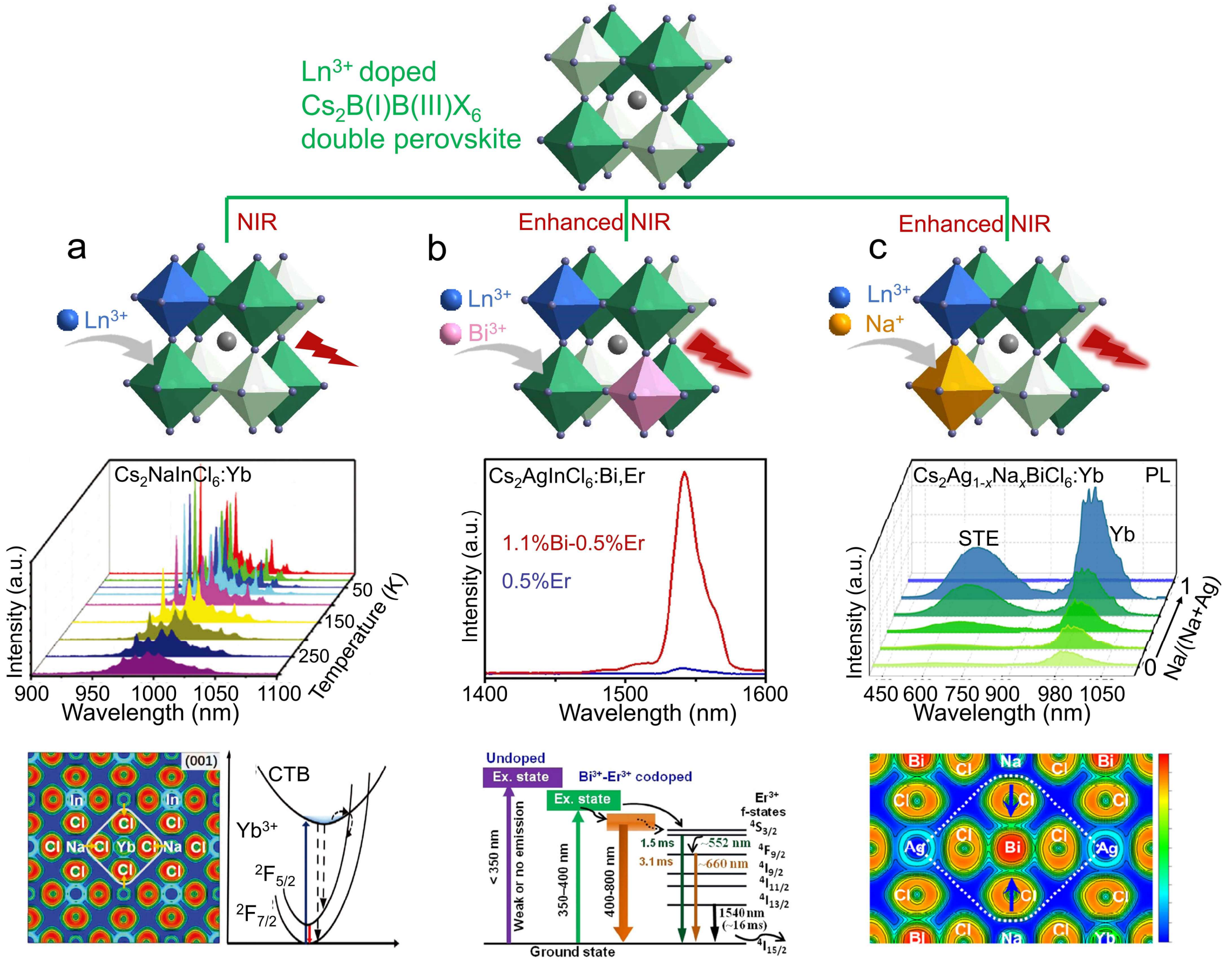

**Figure 5. $Ln^{3+}$ doped double perovskites with NIR emission.** (a) Temperature-dependent PL spectra ($\lambda_{ex}$ = 273 nm), the electron localization function (ELF) analysis and sketch of $Yb^{3+}$ electronic transitions for $Cs_2NaInCl_6:Yb^{3+}$. Reproduced (Adapted) with permission.[63d] Copyright 2022, Wiley. **$Ln^{3+}$ and other ions co-doped double perovskites with enhanced NIR emission.** (b) PL spectra in the NIR region and simplified energy level schemes of $Er^{3+}$ doped and $Bi^{3+}$–$Er^{3+}$-co-doped $Cs_2AgInCl_6$. Reproduced (Adapted) with permission.[24a] Copyright 2020, Wiley. (c) PL spectra ($\lambda_{ex}$ = 360 nm) and ELF analysis of $Na^+$ alloyed $Cs_2AgBiCl_6:Yb^{3+}$. Reproduced (Adapted) with permission.[70] Copyright 2022, Wiley.

It has been reported that $Na^+$ alloying could break the inversion-symmetry-induced parity-forbidden transition of $Cs_2AgInCl_6$ and thus improve its broadband emission.[23a] Based on that, Zhang et al. synthesized $Cs_2Na_{0.2}Ag_{0.8}InCl_6:Yb^{3+}$ single crystals with the NIR PLQY up to 46%.[69] The obtained single crystals showed dual emission in the visible and NIR region activated by STE and $Yb^{3+}$ ions excited at 365 nm. The NIR emission at 996 nm ($^2F_{5/2} \rightarrow {}^2F_{7/2}$) originated from the energy transfer from STE to $Yb^{3+}$. After removing the excitation, a long-lasting afterglow at 996 nm with a duration up to 7200 s was observed in the $Cs_2Na_{0.2}Ag_{0.8}InCl_6$:$Yb^{3+}$ crystals.[69] Chen et al. reported that through $Na^+$ alloying, the NIR PLQYs of $Yb^{3+}$ and $Er^{3+}$ doped $Cs_2Ag_{0.2}Na_{0.8}BiCl_6$ were enhanced 7.3-fold and 362.9-fold compared with those in Na-free $Cs_2AgBiCl_6$, respectively (Figure 5c).[70] The intensity of the PLE monitored at 995 nm and of the PL spectra were greatly improved after $Na^+$ doping (Figure 5c).[70] The electron localization function (ELF) provided evidence that $Na^+$ doping could break the local site symmetry and thus promote the absorption of $Bi^{3+}$ and improve the NIR emission (Figure 5c).[70] A possible mechanism was proposed that included the absorption of the transitions of $Bi^{3+}$ ions and the effective energy transfer process to the STE states and then to $Yb^{3+}$/$Er^{3+}$ ions.[70]

Ln-based double perovskites $A_2B^+B^{3+}X_6$, are obtained when part (or all) of the $B^{3+}$ ions are $Ln^{3+}$.[73] Among them, $Cs_2NaErCl_6$ is the most reported one. Lin et al. synthesized $Cs_2NaErCl_6$ single crystals and alloyed $Yb^{3+}$ into them.[73c] Unalloyed $Cs_2NaErCl_6$ showed visible and NIR PL bands at about 484, 513, 559, 664, 807, 995 and 1542 nm, attributed to the *f*–*f* transitions of $Er^{3+}$, respectively. When alloyed with $Yb^{3+}$, the red emission at 664 nm from $Er^{3+}$ and the NIR emission at 995 nm from $Yb^{3+}$ under 379 nm excitation were both enhanced. The PLQY of an optimized $Cs_2NaEr_{0.4}Yb_{0.6}Cl_6$ sample was 62%.[73c] Han et al. synthesized a series of $Cs_2NaEr_{1-x}B_xCl_6$ (B = In, Sb, or Bi; $x$ = 0, 0.13, and 0.5) NCs.[73b] The $Cs_2NaErCl_6$ NCs exhibited a weak NIR PL peaking at 1543 nm from the $^4I_{13/2} \rightarrow {}^4I_{15/2}$ transition of the $Er^{3+}$, with an average lifetime of 35.7 μs excited at 380 nm.[73b] Wang et al. doped $Mn^{2+}$ into $Cs_2NaBi_{1-x}Er_xCl_6$ to further improve its PLQY.[73a] In the visible region, $Cs_2NaBi_{0.86}Er_{0.14}Cl_6$:$Mn^{2+}$ showed broadband emission from $Mn^{2+}$ and sharp emissions from $Er^{3+}$. In the NIR region, $Cs_2NaBi_{0.86}Er_{0.14}Cl_6$:$Mn^{2+}$ exhibited PL at 1540 nm with increased PLQY to ~14% compared with that of $Cs_2NaBi_{0.86}Er_{0.14}Cl_6$.[73a] The increased PLQY was attributed to the incorporated $Mn^{2+}$ that act as the intermediate bridge between the energy levels of the host and $Er^{3+}$, resulting in a more efficient energy transfer process from the host to $Er^{3+}$. In addition, they also synthesized $Cs_2Na(Bi,Ho)Cl_6$:$Mn^{2+}$ and $Cs_2Na(Bi,Nd)Cl_6$:$Mn^{2+}$ and obtained intense NIR emission from the transitions of $Ho^{3+}$ and $Nd^{3+}$ ions with the PLQYs of ~6% and ~35%, respectively.[73a] Chen et al. prepared a class of lanthanide double perovskite NCs exhibiting emission covering a wide spectral window, from UV-C to NIR (260–995 nm).[74] Among them, ternary alloy $Cs_2NaCe_{0.9}Tb_{0.09}Yb_{0.01}Cl_6$ NCs showed a NIR Yb-related PL at 995 nm with a PLQY of 6%, 30 times higher than that of unalloyed $Cs_2NaYbCl_6$ NCs.

The enhancement was reported to be due to energy transfer from the Ce host to the Yb dopant bridged by the Tb component.[74] More recently, a Ho-based double perovskite, $Cs_2NaHoCl_6$, with a wide direct bandgap of ~5 eV (similar to $Cs_2NaErCl_6$) was reported to have red and NIR emissions at 650, 980, 1200, and 1450 nm under blue light excitation, corresponding to the $^5F_5 \rightarrow {}^5I_8$, $^5F_5 \rightarrow {}^5I_7$, $^5I_6 \rightarrow {}^5I_8$, and $^5F_5 \rightarrow {}^5I_6$ transitions from $Ho^{3+}$ ions.[75] As one of Ln-based double perovskites, $Cs_2NaHoCl_6$ showed limited PL efficiency due to its small absorption cross section and high concentration quenching. Therefore, ions doping/co-doping ($Sb^{3+}$, $Bi^{3+}$ and $Ag^{+}$) in $Cs_2NaHoCl_6$ was exploited to improve the PLQY of its NIR emission up to 17.6%.[75]

### 4.3 $Ln^{3+}$ ions doped into other metal halides

In addition to double perovskites, various groups have investigated other lead-free metal halides with octahedral coordination environment as hosts for $Ln^{3+}$ ions dopants, including 0D Zr-based, 1D Mn-based and 2D Bi-based metal halides.[8d, 24b, 24c] The vacancy-ordered double perovskite $Cs_2ZrCl_6$ has a direct bandgap and features a relatively efficient emission in the visible region. Chen et al. co-doped $Te^{4+}$ and $Ln^{3+}$ ions ($Er^{3+}$, $Nd^{3+}$ and $Yb^{3+}$) into $Cs_2ZrCl_6$ (**Figure 6**a, b), obtaining dual emission bands centered at 575 nm and in the NIR region peaking at 1539, 1073 and 1002 nm, respectively, under 392 nm excitation.[24c] The dual emissions originated from $^1S_0 \rightarrow {}^3P_1$ transition of $Te^{4+}$ and *f–f* transitions of $Ln^{3+}$ ions. The strongest emitting sample had NIR PLQY of 6.1%, arising from the $Er^{3+}$ emission, sensitized by $Te^{4+}$ ions through an energy transfer process from $Te^{4+}$ to $Er^{3+}$ (Figure 6c).[24c] Stable vacancy-ordered quadruple perovskites such as $Cs_4(Mn,Cd)(Bi,Sb)_2Cl_{12}$ could also be doped with $Ln^{3+}$ ions ($Nd^{3+}$, $Dy^{3+}$, $Ho^{3+}$, $Er^{3+}$, $Tm^{3+}$, $Yb^{3+}$) to achieve NIR emissions covering the NIR-I and NIR-II regions.[76] Mn alloying could improve the NIR emissions originating from the $Mn^{2+} \rightarrow Ln^{3+}$ energy transfer process. Also, $Bi^{3+}$ alloying enhanced the emission of $Mn^{2+}$ but quenched the $Ln^{3+}$ emissions due to the increased mismatching between the $^4T_1$ energy level of $Mn^{2+}$ and that of $Ln^{3+}$.[76c]

Cesium manganese bromides can also act as broad visible absorbers for the NIR emission from $Ln^{3+}$ ions through a downshifting process. For example, Jin et al. synthesized $Yb^{3+}$, $Er^{3+}$ and $Ho^{3+}$ doped $CsMnCl_3$ single crystals with visible emission at 655 nm and NIR emissions at 1002 nm for $Yb^{3+}$, 801, 977 and 1538 nm for $Er^{3+}$ and 907 and 1183 nm for $Ho^{3+}$ upon excitation at 528 nm. The NIR emission bands from $Ln^{3+}$ ions were reported to be sensitized by an energy transfer process from $Mn^{2+}$ and the highest PLQY for $Yb^{3+}$ doped $CsMnCl_3$ was 7%. Jalali et al. attempted to dope $Ln^{3+}$ ions into $CsMnBr_3$ and $Cs_3MnBr_5$ NCs and found that only $CsMnBr_3$

NCs could host $Ln^{3+}$ ions (Figure 6d-f).[8d] This is due to the octahedral coordination of the $Mn^{2+}$ sites in $CsMnBr_3$, favorable for hosting $Ln^{3+}$ ions, as opposed to the unfavorable tetrahedral coordination of $Cs_3MnBr_5$. Sharp NIR emission bands at 890 and 1075 nm from $Nd^{3+}$, 1005 nm from $Yb^{3+}$, 1226 and 1489 nm from $Tm^{3+}$, and 1544 nm from $Er^{3+}$ were observed upon excitation at 550 nm. The value of NIR PLQYs was in the range of 0.24–1.1%. The PLE spectra monitored at the NIR emission bands of $CsMnBr_3$:$Ln^{3+}$ NCs evidenced the *d-d* transition of the $Mn^{2+}$ ion in $d^5$ configuration, the same as undoped $CsMnBr_3$, indicating that the NIR emissions from $Ln^{3+}$ was sensitized by the excitation of the host.[8d] The band structures computed by DFT indicated that the $5f_{5/2}$ and $5f_{7/2}$ states of Yb lied deep inside the bandgap of the material (Figure 6f), enabling the NIR emission from the dopant.[8d]

Coming to Bi-based metal halides, $Cs_3Bi_2X_9$ (X = $Cl^-$, $Br^-$, $I^-$) have bandgaps and band emissions that can be tuned in the entire visible range by replacing $Br^-$ with $Cl^-$ or $I^-$.[8f] That makes $Cs_3Bi_2X_9$ an interesting host for $Yb^{3+}$ dopants to elicit a relatively high PLQY arising from quantum cutting. Besides, the $Yb^{3+}$ ion as dopant is isovalent to $Bi^{3+}$ and thus it would tend to preserve the $Cs_3Bi_2Br_9$ structure. Based on that, Eray S. Aydil et al doped $Yb^{3+}$ in $Cs_3Bi_2Br_9$ films, demonstrating double emissions in the visible and NIR regions, under 420 nm and 360 nm excitations, respectively (Figure 6g–i).[24b] The NIR emission of $Yb^{3+}$ with 14.5% PLQY was reported to originate from a possible quantum cutting effect.[24b] Recently, $Er^{3+}$ was doped into $CsCdCl_3$ crystals, leading to two NIR emission bands in the region of 1400–1750 nm, centered at 1534 and 1680 nm, respectively, under the excitation of 800 nm.[77] The two emission bands were related to the $^4I_{13/2} \rightarrow {}^4I_{15/2}$ and $^4S_{3/2} \rightarrow {}^4I_{9/2}$ transition of $Er^{3+}$, respectively.[77]

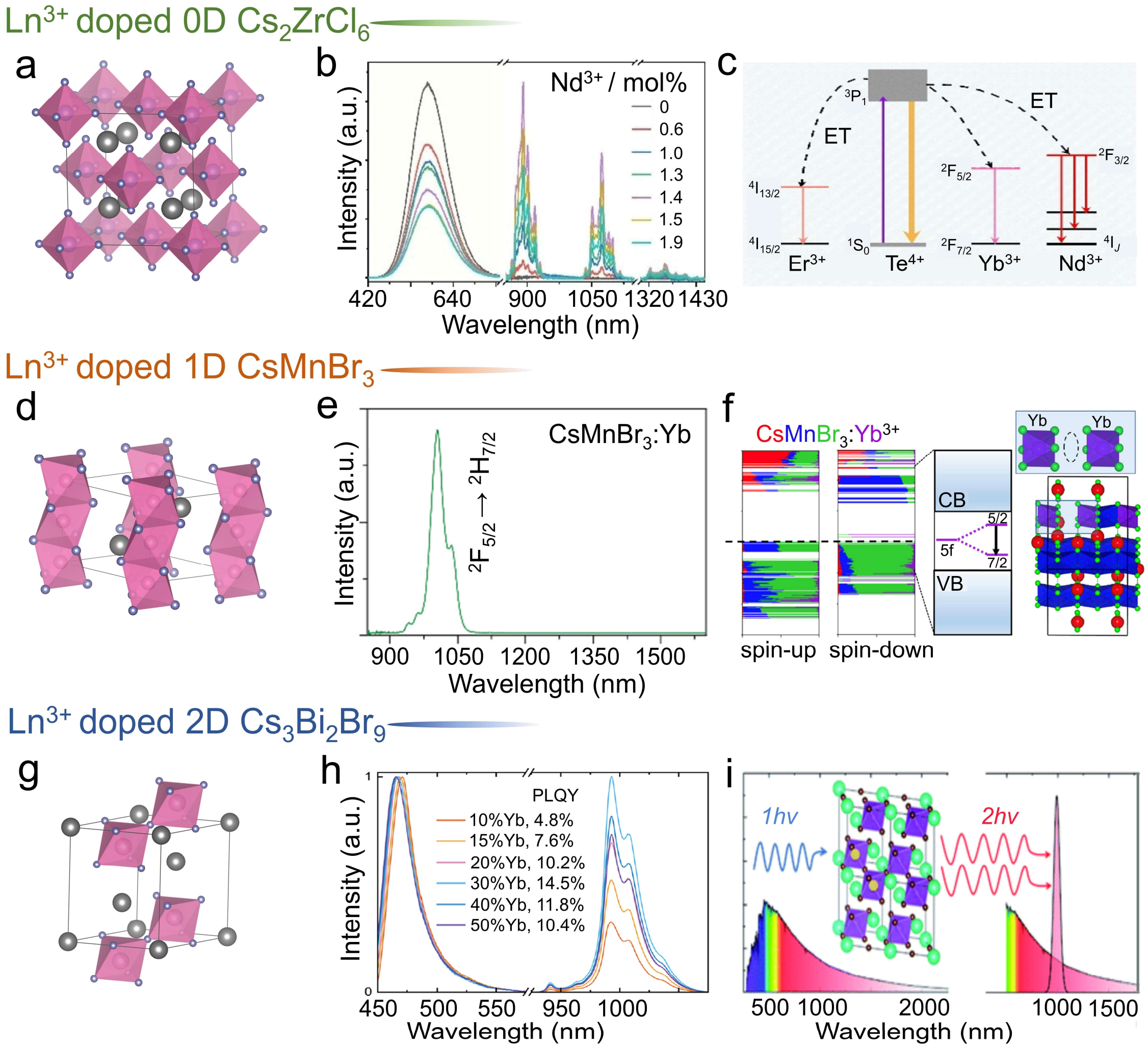

**Figure 6. $Ln^{3+}$ doped 0D $Cs_2ZrCl_6$.** (a) Crystal structure of $Cs_2ZrCl_6$; (b) PL spectra ($\lambda_{ex}$ = 392 nm) of $Cs_2ZrCl_6$:$Te^{4+}$/$Nd^{3+}$ MCs; (c) Energy levels of $Te^{4+}$, $Er^{3+}$, $Nd^{3+}$ and $Yb^{3+}$ in $Cs_2ZrCl_6$ MCs. Reproduced (Adapted) with permission.[24c] Copyright 2022, Wiley. **$Ln^{3+}$ doped 1D $CsMnBr_3$.** (d) Crystal structure $CsMnBr_3$; (e) PL spectra ($\lambda_{ex}$ = 550 nm) in the NR region of $CsMnBr_3$:$Yb^{3+}$ NCs. (f) Energy level diagram of $CsMnBr_3$:$Yb^{3+}$. Reproduced (Adapted) under the terms of the CC-BY Creative Commons Attribution 4.0 International License.[8d] Copyright 2022, The Authors, published by American Chemical Society. **$Ln^{3+}$ doped 2D $Cs_3Bi_2Br_9$.** (g) Crystal structure of $Cs_3Bi_2Br_9$; (h) Normalized visible PL ($\lambda_{ex}$ = 420 nm) and NIR PL ($\lambda_{ex}$ = 360 nm) from $Cs_3Bi_2Br_9$:$Yb^{3+}$ films; (i) Proposed PL mechanism of $Cs_3Bi_2Br_9$:$Yb^{3+}$. Reproduced (Adapted) with permission.[24b] Copyright 2021, Royal Society of Chemistry.

## 5. Metal halides with broadband NIR emission

### 5.1 Self-trapped Exciton recombination

An STE in metal halides can generate NIR luminescence characterized by broadband emissions with a large Stokes shift, as anticipated in the introduction. An STE-related emission is different from the one arising from a permanent lattice defect (for example a vacancy or an anti-site): the STE-related lattice deformation occurs only transiently under light excitation, whereas a permanent defect is inherent of the crystal lattice and generally leads to defect states within the bandgap. The intensity of the luminescence from permanent lattice defects is closely related to the density of such defects. In the process of STE formation, the energy lost by the exciton due to self-trapping is named as the self-trapping energy $E_{st}$, and the increase in energy of the ground state due to the local lattice distortion is defined as the lattice deformation energy $E_d$ (**Figure 7**a). If $E_g$ is the energy of the gap and $E_b$ is the exciton binding energy, then the energy $E_{PL}$ of the STE emission is equal to $E_g - E_b - E_{st} - E_d$.[11a] This formula explains the reason for the large Stokes shifts in STE-related emission. If the material has a reduced bandgap, a large exciton binding energy, a large self-trapping energy, or large lattice deformation energy, then most likely an STE-related emission will be in the NIR. Strong electron–phonon coupling is essential to the formation of STE and can be evaluated by the Huang-Rhys factor, *S*. As shown in the following equation,

$$\mathrm{FWHM}(T) = 2.36\sqrt{S}\hbar\omega_{phonon}\sqrt{\coth\frac{\hbar\omega_{phonon}}{2k_BT}}$$

the Huang-Rhys factor, *S*, has a positive correlation with the full-width-at-half-maximum (FWHM). Thus, a strong electron–phonon coupling leads to a broad emission band for a STE-related emission. The most obvious strategy to achieve NIR emission is to reduce the fundamental bandgap of the metal halide ($E_g$). The $ns^2$ lone pairs of the metal cation can interact with the *p* orbitals of the halogen anions, forming anti-bonding orbitals higher than the VBM and thus reducing the bandgap.[78] Besides, for a given material, the bandgap gets reduced with the increase in connectivity of the octahedra, as a result of increased overlap of the atomic orbitals that participate in forming the band edge states.[79] With increase in connectivity, the VBM and CBM become more dispersive (they have larger band widths), thus leading to a narrower bandgap.[79-80] Therefore, designing metal halides with high connectivity is another effective approach for reducing the bandgap. In addition, the structural symmetry of metal halides in which the fundamental units are multiple polyhedra (such as $[Sb_2X_8]^{2-}$ dimers in $(C_{13}H_{22}N)_2Sb_2Cl_8$) is lower than that of halides in which such units are individual polyhedra (such as $[SbX_4]^-$ in $(C_{16}H_{36}P)SbCl_4$), resulting in greater freedom of structural deformation and larger structural deformation energy ($E_d$).[30e] Therefore, metal halides with lower structural symmetry may lead to a larger Stokes shift, towards the NIR emission region.

Electron–phonon coupling is essential for the formation and emission efficiency of the STE induced NIR emission. On the one hand, strong electron–phonon coupling can facilitate the formation of STE. On the other hand, a too strong coupling can promote nonradiative recombination processes and decreased PL efficiency.[11a] Note that the electron–phonon coupling can be evaluated by the Huang-Rhys factor *S*. A larger *S* value means a stronger electron–phonon coupling. The coordinate difference between free exciton state and STEs, $\Delta Q$, positively correlates with FWHM and *S*. Therefore, an excessively large *S* results in a large $\Delta Q$, thus leading to the crossing of the excited- and ground-state curves (Figure 7b). The crossing of the two curves means that some excited state electrons and holes recombine in a nonradiative way with the phonons emitted, resulting in a reduced PLQY (Figure 7b).[11a, 30g] In this section, we will discuss the designs of STE-induced broadband NIR emission and strategies for modulating the PL efficiency of NIR emission in metal halides.

**Table 2.** Optical parameters of STE based NIR emission in different metal halides

| Material | PL peak (nm) | Stokes shift | FWHM (nm) | PLQY (%) | Ref. |
|---|---|---|---|---|---|
| $Cs_2Ag_{0.05}Na_{0.95}BiCl_6$ single crystal | 700 | 340 nm | 270 | 51 | [30f] |
| $Sn^{2+}$ alloyed $Cs_2ZnBr_4$ phosphor | 700 | 323 nm | 177 | 41 | [30g] |
| $Cs_2WCl_6$ single crystal<br>$Cs_2MoCl_6$ single crystal | ~950<br>~1000 | 550–600 nm | 200-300 | 26 | [30d] |
| $Bmpip_2SnI_4$ phosphor | 730 | 375 nm | / | 75 | [30a] |
| $(TMEDA)SbI_5$ single crystal | 714 | 483 nm | / | / | [30b] |
| $(MTP)_2SbBr_5$ glass | 735 | 285 nm | / | 5.5 | [30c] |
| $(C_{13}H_{22}N)_2Sb_2Cl_8$ single crystal | 865 | 515 nm | 285 | 5 | [30e] |
| $(C_{10}H_{16}N)_2Sb_2Cl_8$ single crystal | 990 | 645 nm | 336 | 3 | [30e] |
| $(C_{16}H_{36}P)SbCl_4$ single crystal | 1070 | 735 nm | 331 | 1 | [30e] |

The STE recombination is typically characterized by large Stokes shifts and broadband emissions (FWHM >100 nm). In proper conditions, STE based NIR emission can be achieved in some metal halides with either low electronic dimensionality (low connectivity of the atomic orbitals) or low structural dimensionality (low connectivity of the polyhedra). The optical parameters of STE based NIR photoluminescence in different metal halides are listed in **Table 2**. For inorganic metal halides, halide double perovskites with 0D electronic dimensionality could generate STE. For example, Wang et al. reported that by introducing $Na^+$ ions, a $Cs_2AgBiCl_6$ single crystal achieved efficient broadband NIR emission from STE (Figure 7c, d).[30f] Under excitation at 365 nm, the $Cs_2Ag_{0.05}Na_{0.95}BiCl_6$ single crystal emitted NIR emission peaking at 700 nm with a PLQY of 51% and FWHM of 270 nm (Figure 7c). The enhanced NIR luminescence originated from weakened electron–phonon interaction, reduced $\Delta Q$ and increased exciton binding energy upon introduction of $Na^+$ (Figure 7d).[30f]

0D metal halides with low crystal dimensionality and soft lattice are more prone to lattice distortion and thus promote the STE formation, and relatively high photoluminescence efficiency due to the strong quantum confinement. One example in this class is represented by $Cs_2SnX_6$, having a so-called vacancy ordered perovskite that features isolated $[SnX_6]^{2-}$ octahedra. In these materials, the broad emission, strongly Stokes-shifted from the absorption, appears to be due to an STE which, for the iodide case, is the NIR. For example, Saparov et al. prepared $Cs_2SnI_6$ films with enhanced stability and a direct band gap of 1.6 eV, exhibiting a NIR PL at ~795 nm.[81] For this composition there is a vast literature on possible morphologies and dimensionalities that can be achieved (bulk crystals, films, NCs). For a more in depth coverage of these materials, see our recent review on low dimensional metal halides.[82] The vacancy ordered perovskite structure is not only restricted to the $ASnX_6$ case but can be obtained from various other $M^{4+}$ cations. In some cases, the emission from these materials is in the NIR. One interesting case is that $Cs_2PdBr_6$, seems to present a direct bandgap, as reported by Sakai et al. and Zhou et al.[83] $Cs_2PdBr_6$ single crystals showed PL emission at 772 nm, apparently from the band-to-band transition, while their NC counterpart showed a blue-shift PL emission at ~735 nm, allegedly due to the quantum confinement effect, although quantum confinement effects should not be seen in such 0D structures. Overall, the properties of this material need to be carefully validated by future studies. Another vacancy ordered structure with NIR emission is represented by $Cs_2MoCl_6$ and $Cs_2WCl_6$, which were recently synthesized as single crystals.[30d] With ultraviolet or blue excitation, $Cs_2MoCl_6$ and $Cs_2WCl_6$ single crystals emit light peaking at ~950 nm and ~1000 nm, respectively, with a PLQY of ~26%. The large Stokes shift (550–600 nm) and the broad FWHM (200–300 nm) indicated features typical of a STE.[30d] The realm of 0D metal halides is quite vast and goes well beyond the vacancy ordered perovskites, but only a few cases have emerged so far of NIR emission. As an example, Wang et al. incorporated $Sn^{2+}$ cations into 0D $Cs_2ZnBr_4$ crystals and obtained a broadband NIR emission (**Table 2**).[30g] The Huang-Rhys factor $S$ decreased from 51.81 ($Cs_2ZnBr_4$) to 12.65 ($Cs_2ZnBr_4$:$Sn^{2+}$). The suitable $S$ value after $Sn^{2+}$ alloying not only reduces nonradiative recombination processes due to the strong coupling effect but also ensures the formation of STEs.[30g]

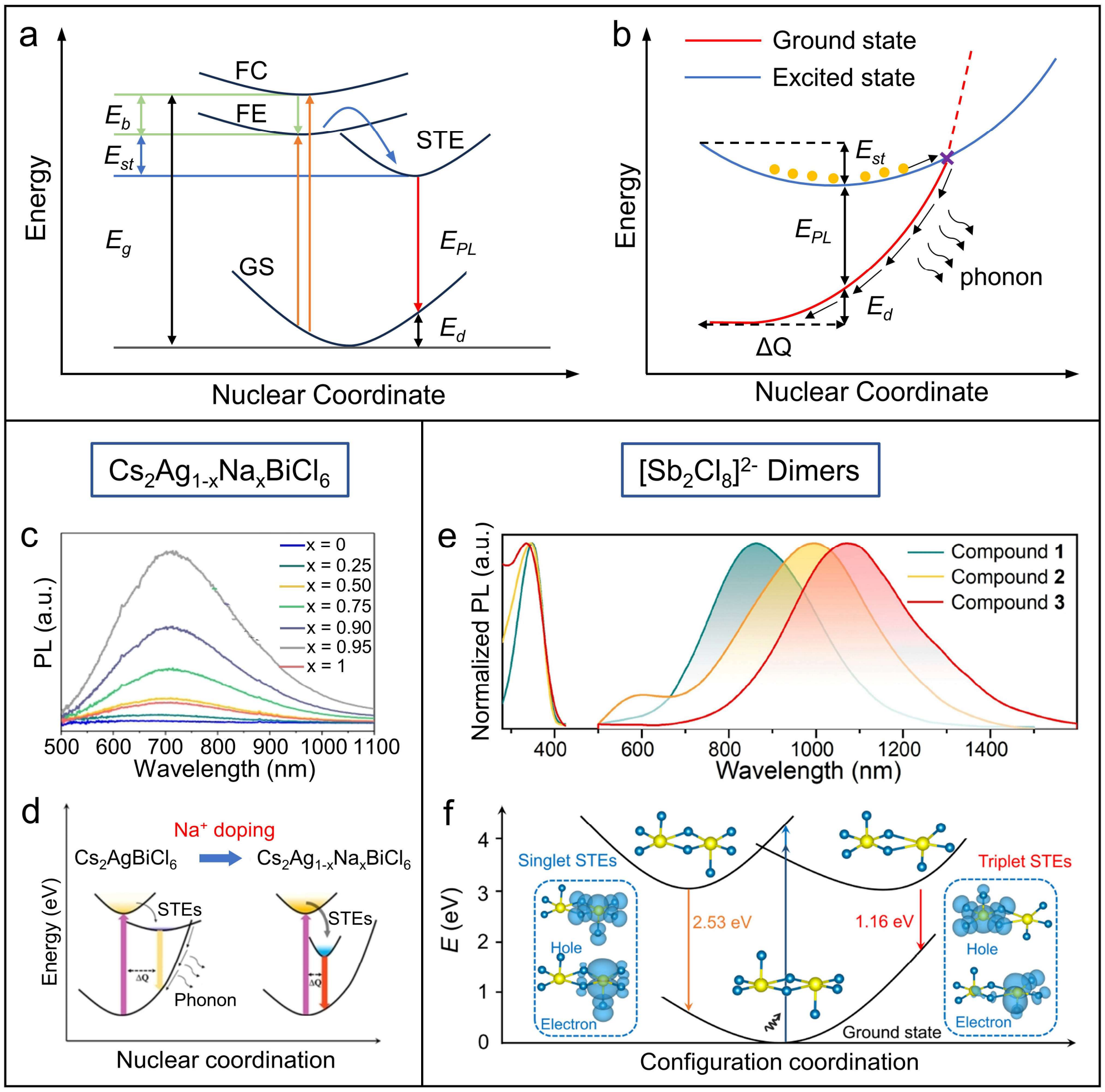

**Figure 7.** (a) Sketch of the energy level structure of STE (FC: free carrier state; FE: free exciton state; GS: ground state). (b) Sketch of the nonradiative recombination process with over large S. Reproduced (Adapted) with permission.[11a] Copyright 2019, American Chemical Society. (c) PL spectra of $Cs_2Ag_{1-x}Na_xBiCl_6$. (d) Schematic of the photophysical processes in $Cs_2AgBiCl_6$ and $Cs_2Ag_{0.05}Na_{0.95}BiCl_6$. Reproduced (Adapted) with permission.[30f] Copyright 2022, Wiley. (e) Normalized PL and PLE spectra of compound **1** ($(C_{13}H_{22}N)_2Sb_2Cl_8$), **2** ($(C_{10}H_{16}N)_2Sb_2Cl_8$), and **3** ($(C_{16}H_{36}P)SbCl_4$). (f) Configuration coordinate diagram for singlet and triplet STEs emissions of compound **2**. Reproduced (Adapted) with permission.[30e] Copyright 2022, Wiley.

Low dimensional organic-inorganic hybrid metal halides consist of organic A-site cations and isolated metal halide polyhedrons. Until now, only a few low-dimensional hybrid metal halides have been reported to show NIR emissions, mainly in the region of 700–1000 nm (**Table 2**). Kovalenko et al. used 1-butyl-1-methylpiperidinium cations ($C_{10}H_{22}N$, Bmpip) to obtain 0D metal halides $Bmpip_2SnI_4$, emitting NIR light at 730 nm under excitation at 355 nm with PLQY of 35%.[30a] Xu et al. synthesized a $(MTP)_2SbBr_5$ glass with blue-light (450 nm) excited NIR emission at 735 nm. The NIR PL exhibited a large Stokes shift (285 nm) with a PLQY of 5.5%.[30c] Zhao et al. reported a NIR emitting 1D hybrid metal halide $(TMEDA)BiI_5$. The broadband NIR emission at 805 nm, excited by 322 nm, showed a large Stokes shift of 483 nm, pointing to a STE related mechanism.[30b] Xia et al. designed a series of NIR-emitting 0D hybrid metal halides composed of $[Sb_2Cl_8]^{2-}$ dimers with low structural symmetry and large freedom for structural distortion.[30e] They chose $[C_{13}H_{22}N]^+$, $[C_{10}H_{16}N]^+$ and $[C_{16}H_{36}P]^+$ as the organic cations, leading to a gradual increase in the extent of structural distortion for the $[Sb_2Cl_8]^{2-}$ dimers. The calculated distortion indexes of the $[Sb_2Cl_8]^{2-}$ dimers in $(C_{13}H_{22}N)_2Sb_2Cl_8$, $(C_{10}H_{16}N)_2Sb_2Cl_8$ and $(C_{16}H_{36}P)SbCl_4$ are 0.065, 0.078 and 0.084, respectively. The three 0D metal halides had NIR broadband emissions at 865 nm, 990 nm and 1070 nm when excited by 350 nm, 345 nm and 335 nm light, with PLQY of 5%, 3% and 1%, respectively (Figure 7e).[30e] The two emission peaks of $(C_{10}H_{16}N)_2Sb_2Cl_8$ originated from singlet and triplet STEs (Figure 7f).[30e]

### 5.2 Specific Ions doped into metal halides

Some specific ions doping ($Bi^{3+}$, $Bi^+$, $Sb^{3+}$ and $Cr^{3+}$) can induce broadband NIR emission and the related optical parameters are listed in **Table 3**. In this section, we will discuss the relevant studies, PL mechanism and designs for ions doping induced broadband NIR emission in metal halides.

**Table 3.** Optical parameters of specific dopants induced NIR emission in metal halides

| Dopant | Host | PL mechanism | PL peak (nm) | FWHM (nm) | PLQY (%) | Ref. |
|---|---|---|---|---|---|---|
| $Bi^{3+}$ | $MAPbBr_3$ crystal | Dopant-induced defect | 1080 | ~400 | / | [84] |
| | $MAPbI_3$ film | | 1140 | 380 | / | [37a] |
| | $MAPbI_3$ film | | 1145 | 375 | / | [37d] |
| | $CsPbI_3$ NCs | | 1145 | ~400 | 7 | [37c] |
| | $CsPbI_3$ powders | | 1185 | 175 | / | [37b] |
| $Bi^+$ | $ABCl_3$ (A = $K^+$, $Rb^+$, $Cs^+$, Cs; B = $Mg^{2+}$, $Cd^{2+}$) | $^3P_1 \rightarrow {}^3P_0$ | 905-1015 | 117-157 | / | [35b] |
| | $Cs_2ZnCl_4$ crystal | | 745 | 175 | 69.9 | [39a] |
| $Sb^{3+}$ | $Cs_2SnCl_4Br_2$ crystal | $^3P_1 \rightarrow {}^1S_0$ | 710 | 164 | 8 | [39b] |
| | $Rb_2InBr_5 \cdot H_2O$ crystal | | 750 | / | 20.8 | [38] |
| | $(NH_4)_2SnCl_6$ microcrystal | Sb-Sn mixed STE | 734 | / | 23 | [39c] |
| | $Cs_2NaInCl_6$ crystal | | ~880 (17 K) | / | / | [31] |
| | $Cs_2NaY(Cl,Br)_6$ crystal | | 910, 975 (17 K) | / | / | [31] |
| $Cr^{3+}$ | $Cs_2NaScCl_6$ crystal | $^4T_2 \rightarrow {}^4A_2$ | 900 | / | 0.7 | [85] |
| | $Cs_2AgInCl_6$ microcrystal | | 1010 | 180 | 22 | [32a] |
| | $Cs_2(Ag,Na)InCl_6$ NCs | | 958-998 | 165-195 | 5.8-19.7 | [32c] |

Bismuth doping has been reported to activate NIR emission attributed to diverse oxidation states: $Bi^+$ and $Bi^{3+}$.[36, 37b] Bismuth ions are characterized by multiplex electronic structures and they could exhibit PL in the visible or NIR spectral regions.[36] Generally, the emission bands of $Bi^+$, $Bi^{2+}$, and $Bi^{3+}$ are much broader than those of *f*–*f* transitions of $Ln^{3+}$. Among the oxidation states, $Bi^{3+}$ is the most common and stable ionic form with the outermost $6s^2$ electronic configuration. $Bi^{3+}$ ions have been introduced in lead-based perovskites leading to broadband NIR PL. For example, after $Bi^{3+}$ doping, $MAPbBr_3$ crystals evidenced an additional PL peak at 1080 nm apart from their intrinsic PL in the visible region.[84] The NIR emission of $Bi^{3+}$ doped $MAPbBr_3$ originated from a possible energy transfer process from the host to the radiative states induced by $Bi^{3+}$ dopants. However, the exact origin of NIR emission from $Bi^{3+}$ doping needs further investigation.[84] Song et al. reported that doping $Bi^{3+}$ into $MAPbI_3$ films provided ultra-broadband NIR PL peaking at ~1140 nm in the range of 850–1600 nm, apart from the intrinsic PL centered at 782 nm.[37a, 37d] The generated NIR PL was attributed to the induced structural defects after Bi-doping.[37d] $Bi^{3+}$ ions have also been doped in $CsPbI_3$ bulk crystals and NCs, leading to the NIR PL peaking at 1185 nm and 1145 nm, respectively.[37b, 37c] To study the mechanism of NIR emission, $Te^{4+}$ was also doped into $CsPbI_3$, and this elicited a similar NIR PL. This result supports the interpretation that the NIR PL was not from Bi ions, but from the dopant-induced defects with a relatively low PLQY (~7% for $CsPbI_3$:$Bi^{3+}$ NCs), which could be created either by $Bi^{3+}$ or by $Te^{4+}$ doping.[37b, 37c] Differently from $Bi^{3+}$, the electronic configuration of the $Bi^+$ ion is $6p^2$, with $^3P_0$ as the ground state and $^3P_1$ and $^3P_2$ as the excited states (**Figure 8**a). $A_1' \rightarrow E'$ is the allowed transition, $A_1' \rightarrow E''$ and $A_1' \rightarrow A_1'$ is the allowed transition.[35a, 36] The $^3P_1 \rightarrow {}^3P_0$ transition in the $Bi^+$ ions results in a broadband NIR emission. Compared with $Bi^{3+}$, $Bi^+$ is relatively unstable and can only exist in some specific materials, such as molten salts, glass, and zeolites.[36] Since 2015, $Bi^+$ ions have been successfully incorporated in some metal halide crystals, such as $ABCl_3$ ($A = K^+$, $Rb^+$, $Cs^+$; B =

$Mg^{2+}$, $Cd^{2+}$), leading to a similar long-lived broadband NIR emission at ~1000 nm under excitation in the orange-red region.[34-35, 86] In those metal halides, the large alkali ions ($K^+$, $Rb^+$, $Cs^+$) can provide substitution sites for $Bi^+$ ions. [35b] The dopant $Bi^+$ ions are reported to be formed via a synproportionation reaction from bismuth chloride ($BiCl_3$) and Bi metal: $2Bi + BiCl_3 \leftrightarrow 3Bi^+ + 3Cl^-$. The Lewis acidic chlorides $MgCl_2$ and $CdCl_2$ can facilitate the formation of $Bi^+$ by scavenging the $Cl^-$ ions according to the process: $Cl^- + Mg(Cd)Cl_2 \leftrightarrow Mg(Cd)Cl_3^-$. As one example from these works, Sulimov et al. reported $Bi^+$ doped $CsCdCl_3$ single crystals with NIR emission at 980 nm at room temperature from the ${}^3P_1 \rightarrow {}^3P_0$ transition of $Bi^+$ (Figure 8b, c).[34]

Similar to $Bi^{3+}$, $Sb^{3+}$ ions with $5s^2$ as the outermost electronic configuration were used as dopants to induce NIR PL with broadband emission and large Stokes shift (Figure 8d). Due to the electron-interaction and spin-orbit coupling, the excited state splits into four energy states, ${}^1P_1$, ${}^3P_0$, ${}^3P_1$ and ${}^3P_2$. Generally, the ${}^1S_0 \rightarrow {}^3P_0$ or ${}^1S_0 \rightarrow {}^3P_2$ transitions are forbidden, the ${}^1S_0 \rightarrow {}^3P_1$ transition is allowed and the ${}^1S_0 \rightarrow {}^1P_1$ one is parity-allowed.[78, 87] Four examples of $Sb^{3+}$ doping induced NIR emission peaking at 700–800 nm were reported with different PL mechanisms.[38-39] The hosts of all these cases are 0D metal halides, in which Sb–Cl polyhedra (tetrahedra, pentahedra and octahedra) are formed. The excitons are strongly confined in these isolated polyhedra, improving the exciton recombination and resulting in relatively high PLQYs.[88] Xia et al. reported NIR-emitting $Sb^{3+}$ doped $Cs_2ZnCl_4$ crystals (Figure 8e, f).[39a] Excited by 316 nm, $Cs_2ZnCl_4$:$Sb^{3+}$ exhibited a broadband NIR PL peaking at 745 nm with the PLQY of 69.9% and FWHM of 175 nm (Figure 8f). The NIR emission of $Cs_2ZnCl_4$:$Sb^{3+}$ was ascribed to the triplet STE emission (${}^3P_1 \rightarrow {}^1S_0$) of the disphenoidal $[SbCl_4]^-$ polyhedra.[39a] Hu et al. reported $Sb^{3+}$ doped $Cs_2SnCl_4Br_2$ with broadband NIR PL peaking at 710 nm with a large FWHM of 164 nm and PLQY of ~8%.[39b] The NIR emission originated from the ${}^3P_1 \rightarrow {}^1S_0$ allowed transitions of $Sb^{3+}$ ions.[39b] Zou et al. doped $Sb^{3+}$ into $(NH_4)_2SnCl_6$ with PL emission at 590 nm and 734 nm, attributed to $SbCl_5$ STE and Sb–Sn mixed STE, respectively. The PLQY of NIR emission at 734 nm is 23%.[39c] In this area, further studies will likely focus on the mechanism of $Sb^{3+}$ induced NIR emission and the correlations between the electronic structure of the $Sb^{3+}$ related polyhedra and their emission properties, also with the aim to additional suitable matrices in which $Sb^{3+}$ doping can elicit NIR emission.

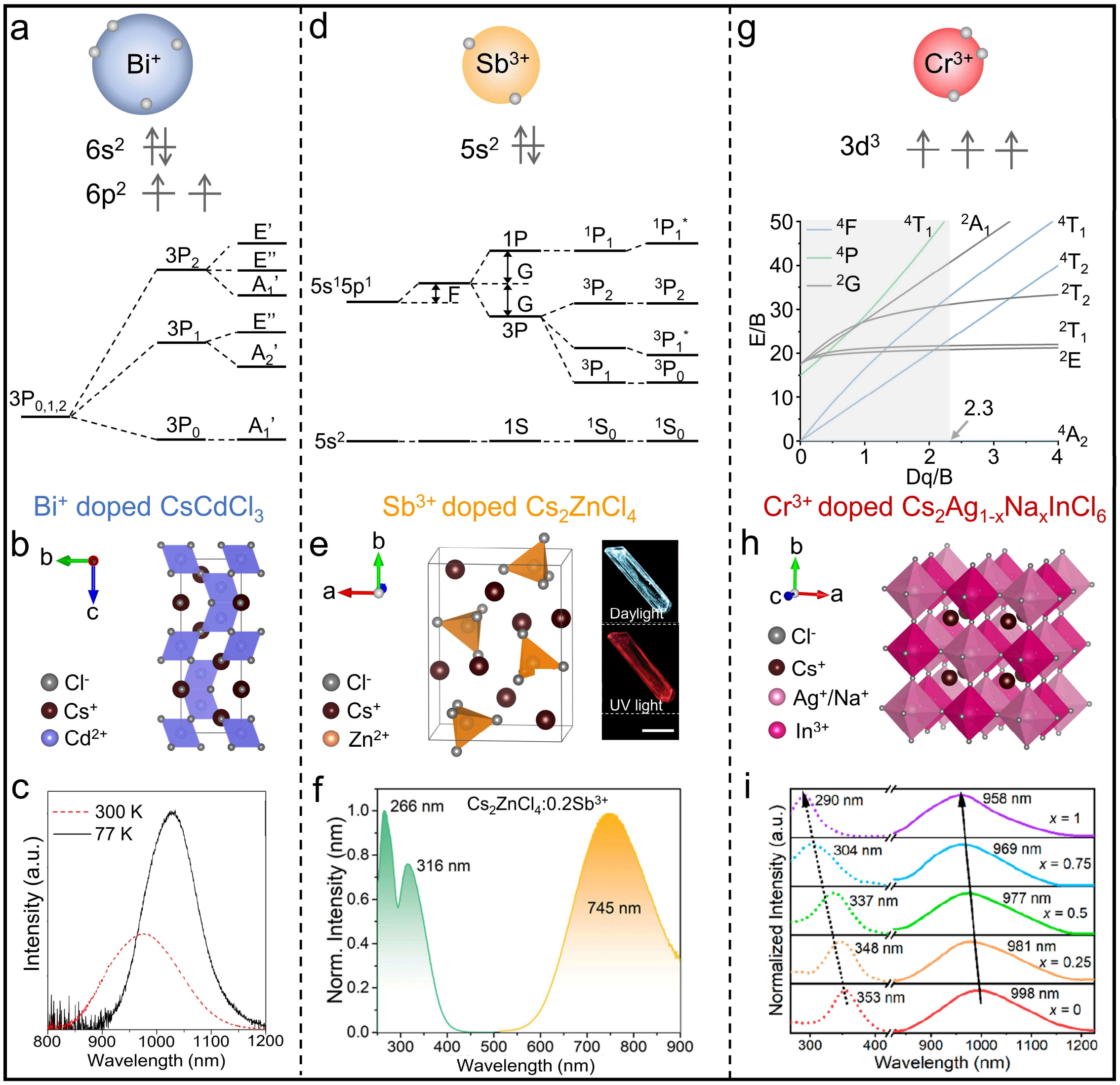

**Figure 8. Bi$^+$ doping induced NIR emission:** (a) Energy level diagram of Bi$^+$. (b-c) Crystal structure and NIR PL spectra of $CsCdCl_3$:Bi$^+$ crystal. Reproduced (Adapted) with permission.[34] Copyright 2014, Elsevier. **Sb$^{3+}$ doping induced NIR emission:** (d) Energy level diagram of Sb$^{3+}$. (e) Crystal structure and sample photographs of $Cs_2ZnCl_4$:Sb$^{3+}$ crystal. (f) Normalized PL and PLE spectra. Reproduced (Adapted) with permission.[39a] Copyright 2021, Wiley. **Cr$^{3+}$ doping induced NIR emission:** (g) Tanabe-Sugano energy level diagram for Cr$^{3+}$. (h-i) The structure, PLE and PL spectra of $Cs_2Ag_{1-x}Na_xInCl_6$:Cr$^{3+}$ NCs. Reproduced (Adapted) with permission.[32c] Copyright 2022, American Chemical Society.

Cr$^{3+}$ ions could also be used as a broadband NIR luminescent activators, especially in phosphors.[33, 89] The emission from Cr$^{3+}$ is strongly affected by the crystal field.[89-90] Figure 8g shows the Tanabe-Sugano diagram of Cr$^{3+}$ ions with 3d$^3$ configuration in the octahedral crystal field. The crystal field can be evaluated by the value of $D_q/B$ ($D_q$: the crystal field strength; B:

the Racah parameter representing the interelectron interaction). As shown in Figure 8g, when the $Cr^{3+}$ ion is in a strong crystal field ($D_q/B > 2.3$), $^2E$ is the lowest excited state. Due to the weak coupling between the $^2E$ level and the host lattice, the narrow-band emission from $^2E \rightarrow {}^4A_2$ spin-forbidden transition takes place. On the other hand, in a weak crystal field ($D_q/B < 2.3$), $^4T_2$ is the lowest excited state and thus the spin-allowed $^4T_2 \rightarrow {}^4A_2$ transition with lower energy occurs. Because of the strong electron–phonon interaction effect for the $^4T_2$ state, $Cr^{3+}$ presents a broadband emission at longer wavelength (650–1350 nm).[91] Based on these considerations, to produce a broadband NIR emission from $Cr^{3+}$ ions an inorganic host with a weak crystal field and appropriate bandgap is needed. Until now, $Cr^{3+}$ ions were only doped/alloyed into double halide perovskites, including $Cs_2NaInCl_6$, $Cs_2NaY(Cl,Br)_6$, $Cs_2NaScCl_6$ and $Cs_2AgInCl_6$, obtaining NIR emission.[31-32, 85] They possess octahedral coordination and the calculated crystal field ($D_q/B$) of $Cs_2NaInCl_6$, $Cs_2NaScCl_6$ and $Cs_2AgInCl_6$ are reported to be 2.28, 2.22 and 2.15–2.25, respectively, meeting the weak crystal field requirement. As an example, in 1986, Knochenmuss et al. doped $Cr^{3+}$ into $Cs_2NaInCl_6$ and $Cs_2NaYX_6$ (X = Cl, Br) single crystals, obtaining broadband NIR emissions centered at ~880, ~910 and ~975 nm at low temperature (<17 K).[31] In 2001, Güdel et al. reported $Cs_2NaScCl_6:Cr^{3+}$ single crystals with a broadband NIR PL at 900 nm excited by 540 nm at room temperature with a PLQY of ~0.7%.[85] In recent years, efforts to obtain efficient NIR emission from $Cr^{3+}$ ions have intensified. Liu et al doped $Cr^{3+}$ into $Cs_2AgInCl_6$ bulk crystals and obtained a broadband NIR emission centered at 1010 nm with a FWHM of 180 nm excited at 353 nm.[32a] The NIR PL was assigned to the spin-allowed $^4T_2 \rightarrow {}^4A_2$ transition of $Cr^{3+}$ ions in octahedral coordination. The PLQY of the NIR emission at 1010 nm under 760 nm excitation was reported to be ~22%.[32a] Following that work, Xia et al. reported the synthesis of $Cr^{3+}$ doped $Cs_2AgInCl_6$ NCs with similar NIR emission.[32c] The NIR PL could be tuned from 998 to 958 nm by replacing $Ag^+$ with $Na^+$, with the spectral tunability being attributed to the gradually enhanced crystal field around $Cr^{3+}$ (Figure 8h, i). The highest NIR PLQY was 19.7%, obtained from a sample of $Cs_2NaInCl_6:Cr^{3+}$ NCs.[32c] When $Yb^{3+}$ was added as a co-dopant, the NIR PLQY of the $Cs_2AgInCl_6:Cr^{3+}$ single crystals was further increased from 22% to 45%.[32b] The $Cs_2AgInCl_6:Cr^{3+},Yb^{3+}$ single crystals exhibited a broadband NIR emission peaking at 1000 nm with the FWHM of 188 nm excited by 365 nm, from the $^4T_2 \rightarrow {}^4A_2$ transition of $Cr^{3+}$. The enhanced PLQY of the $Cs_2AgInCl_6:Cr^{3+},Yb^{3+}$ was ascribed to the $Yb^{3+}$-induced reduced non-radiative recombination and promoted energy transfer from STE to $Cr^{3+}$ after $Yb^{3+}$ doping.[32b] Besides, Nag et al. co-doped $Bi^{3+}$ and $Cr^{3+}$ into $Cs_2Ag_{0.6}Na_{0.4}InCl_6$.[92] Benefiting from the $6s^2 \rightarrow 6s^16p^1$ excitation of $Bi^{3+}$ and energy transfer from $Bi^{3+}$ to $Cr^{3+}$ *d*-electron, a broad NIR emission with peak at 1000 nm and 16% PLQY was achieved, also from the spin-allowed $^4T_2 \rightarrow {}^4A_2$ transition of $Cr^{3+}$.[92]

## 6. Applications of NIR-emitting metal halides

### 6.1 Light-emitting diodes

NIR light-emitting diodes (LEDs) are widely utilized in telecommunication, medical imaging, sensing and security cameras.[93] Traditional NIR LEDs based on GaAs, AlGaAs, and AlAs materials are fabricated under demanding conditions such as high-temperature, ultra-high vacuum, or expensive precursors. NIR emitting metal halides instead are characterized by low-cost synthesis and efficient emission, making them an interesting candidate for the fabrication of a new generation of NIR LEDs. An infrared LED using $MAPbI_{3-x}Cl_x$ as an emitter was firstly reported by Richard H. Friend et al. in 2014.[16a] The radiance of the device was 13.2 W $sr^{-1}$ $m^{-2}$ at a current density of 363 mA $cm^{-2}$ and an external quantum efficiency (EQE) of 0.76%.[16a] Since then, the EQE of NIR LEDs with metal halides has improved from ~0.7% to ~23.8% and the functional lifetime has reached over 11,539 h in less than 10 years of development.[16d, 94] A summary of the performance of NIR LEDs based on metal halides including EQE, operation stability and maximum radiance is reported in **Table 4**.

**Table 4.** Summary of device performance for NIR LEDs based on different metal halides.

| Material | EQE (%) | EL peak (nm) | Operation stability | Radiance (W $sr^{-1}$ $m^{-2}$) | Year | Ref. |
|---|---|---|---|---|---|---|
| $MAPbI_{3-x}Cl_x$ | 0.76 | 754 | / | 13.2 | 2014 | [16a] |
| $MASnI_3$ | 0.72 | 945 | / | 3.4 | 2016 | [18] |
| $MAPbI_3$ | 0.76 | 760 | / | 12.31 | 2016 | [95] |
| | 10.4 | 748 | / | / | 2017 | [96] |
| $MAPbI_3$:$Bi^{3+}$ | / | 1100 (84 K) | / | / | 2016 | [37a] |
| $Cs_{10}(MA_{0.17}FA_{0.83})_{(100-x)}Pb(Br_xI_{1-x})_3$ | 9.23 | 750 | / | 26.19 | 2017 | [97] |
| $FAPbI_3$ | 5 | 801 | / | 206.7 | 2018 | [98] |
| | 20.7 | 803 | 20 h @ 100 mA $cm^{-2}$ ($T_{50}$) | 390 | 2018 | [5a] |
| | 5.2 | 776 | 100 h @ 25 mA $cm^{-2}$ ($T_{50}$) | 88.5 | 2019 | [99] |
| | 20.2 | 799 | 20 h @ 57 mA $cm^{-2}$ ($T_{80}$) | 170 | 2019 | [16c] |
| | 21.6 | 800 | 20 h @ 25 mA $cm^{-2}$ ($T_{50}$) | 308 | 2019 | [5b] |
| | 5.7 | 799 | / | 2.8 | 2020 | [100] |
| | 18.6 | 802 | 682 h @ 20 mA $cm^{-2}$ ($T_{50}$) | 200 | 2021 | [101] |
| | 5.25 | 764 | / | / | 2021 | [102] |
| | 15.4 | 772 | 2, 8.5, 15 min @ 60, 30, 10 mW $m^{-2}$ ($T_{50}$) | / | 2022 | [103] |
| | 22.8 | 803 | 11,539 h @ 5 mA $cm^{-2}$, 32,675 h @ 3.2 mA $cm^{-2}$ ($T_{50}$) | 278.9 | 2022 | [94b] |
| | 23.8 | 800 | 32 h @ 100 mA $cm^{-2}$ ($T_{50}$) | 107 | 2023 | [16d] |
| $FAPbI_3$:$Sn^{2+}$ | 1.7 | 798 | 35 s @ 4.0 V ($T_{50}$) | / | 2019 | [104] |
| $Cs_xFA_{1-x}Pb(Br_{1-y}I_y)_3$ | 5.9 | 735 | 30 s @ 4.5 V ($T_{50}$) | 3.9 | 2018 | [48] |
| FPMAI-$MAPb_{0.6}Sn_{0.4}I_3$ | 5 | 917 | / | 2.7 | 2019 | [105] |
| $CsPbCl_3$:$Yb^{3+}$ | 5.9 | 984 | 58 h @ 0.827 mA $cm^{-2}$ ($T_{50}$) | / | 2020 | [106] |
| $CsPb(Cl_{1-x}Br_x)_3$:$Yb^{3+}$ | 7.7 | 990 | 10 min @ 0.1 mA $cm^{-2}$ ($T_{50}$) | / | 2023 | [107] |
| | 5.4 | 932 | 23.6 h@100 mA cm$^{-2}$ ($T_{50}$) | 162 | 2021 | [108] |
| $CsSnI_3$ | 6.6 | 935 and 951 | 17 h@20 mA cm$^{-2}$ ($T_{50}$) | / | 2023 | [109] |
| | ~2.63 | 948 | 39.5 h@100 mA cm$^{-2}$ ($T_{50}$) | 226 | 2024 | [110] |
| KI-doped $MAPb_{0.8}Sn_{0.2}I_3$ | 9.6 | 868 | / | 0.025 | 2022 | [111] |

First of all, the device structure is crucial to obtain high-quality NIR LED. Depending on different emitting layers (EML), the device is made into conventional architectures or inverted architectures. Conventional or inverted structures consist of anode/hole injecting layer (HIL)/hole transport layer (HTL)/EML/electron transport layer (ETL)/cathode.[112] In the structure, the electrons and holes are injected into the EML layer from cathode/ETL and anode/HTL, where they recombine radiatively. The presence of ETL and HTL enable charges to be injected into EML more effectively. Also, the HTL confines electrons in the EML, while ETL facilitates the electron injection and blocks holes within the EML. The ETLs employed in the NIR device include polyethylenimine ethoxylated (PEIE)-modified ZnO, aluminum zinc oxide (AZO), poly(9,9'-dioctylfluorene) (F8), 1,3,5-tris(1-phenyl-1H-benzimidazol-2-yl)benzene (TPBi), 4,6-Bis(3,5-di(pyridin-3-yl)phenyl)-2-methylpyrimidine (B3PyMPM) and 3',3''',3'''''-(1,3,5-triazine-2,4,6-triyl)tris(([1,1'''-biphenyl]-3-carb) (CN-T2T); and the HTLs can be F8, poly(9,9-dioctyl-fluorene-co-N-(4-butylphenyl)diphenylamine) (TFB), poly[N,N′-bis(4-butylphenyl)-N,N′-bis(phenyl)-benzidine] (Poly-TPD), poly(3,4-ethylenedioxythiophene):polystyrenesulfonate (PEDOT:PSS), poly(Nvinylcarbazole) (PVK) and 9,9-Bis [4-[(4-ethenylphenyl)methoxy]phenyl]-N2,N7-di-1-naphthalenyl-N2,N7-diphenyl-9H-fluorene-2,7-diamine (VB-FPND). The typical structures of NIR LED devices are shown in **Table 5**. In the inverted architectures, ZnO-PEIE is often used as the ETL and poly-TPD or TFB as the HTL. A thin PEIE layer is fabricated onto the ZnO layer to improve the wettability of ZnO and obtain a high coverage of EML. Compared to TFB, Poly-TPD was reported to be a better HTL due to the good energy level alignment for hole injection.[16c] When using poly-TPD as the HTL, there is an enhanced hole-injection efficiency, thus ensuring a balanced supply of holes for radiative recombination with electrons and achieving higher $FAPbI_3$-based device performance.[16c] In the conventional structure, PEDOT:PSS, Poly-TPD or the combination of the organic layers are reported as the HTL and TPBi is often chosen as the ETL. Tseng et al. compared the devices with three organic materials, including PVK, poly-TPD, and VB-FPND as the HTL, respectively.[103] Here, VB-FPND showed the lowest contact angle at the interface, which can improve NCs coverage and achieve higher-efficiency in $FAPbI_3$ NC-based devices. The good adhesion of PEAI-modified $FAPbI_3$ NCs on VB-FPND is attributed to the higher positive zeta potential of VB-FPND, which is beneficial to the coverage of negatively charged NC surface.[103]

**Table 5.** Some typical examples of NIR LED devices structures.

| Inverted Architectures | | | | | | | |
|---|---|---|---|---|---|---|---|
| Year | Cathode | EIL | ETL | Perovskites | HTL | Anode | Ref. |
| 2014 | ITO | $TiO_2$ | | $MAPbI_{3-x}Cl_x$ | F8 | $MoO_3$/Ag | [16a] |
| 2018 | ITO | | ZnO-PEIE | Organic layer /$FAPbI_3$ | TFB | $MoO_x$/Au | [5a] |
| 2019 | ITO | | ZnO-PEIE | $FAPbI_3$ | TFB | $MoO_x$/Au | [5b] |
| | ITO | | ZnO-PEIE | $FAPbI_3$ | Poly-TPD | $MoO_3$/Al | [16c] |
| 2020 | ITO | | AZO-PEIE | $FAPbI_3$ | Poly-TPD | Al/ITO/Ag/ITO | [100] |
| 2021 | ITO | | ZnO-PEIE | $FAPbI_3$ | Poly-TFB | Au | [101] |
| 2022 | ITO | | ZnO-PEIE | SFB10-stabilized $FAPbI_3$ | TFB | $MoO_x$/Au | [94b] |
| | ITO | | ZnO-PEIE | $Cs_xFA_{1-x}PbI_3$ | TFB | $MoO_3$/Au | [5c] |
| 2023 | ITO | | ZnO-PEIE | $FAPbI_3$ | Poly-TPD | $MoO_x$/Au | [16d] |
| Conventional Architectures | | | | | | | |
| Year | Anode | HIL | HTL | Perovskites | ETL | Cathode | Ref. |
| 2016 | ITO | PEDOT:PSS | | $MASn(Br_{1-x}I_x)_3$ | F8 | Ca/Ag | [18] |
| 2017 | ITO | | Poly-TPD | BAI:$MAPbI_3$ | TPBi | LiF/Al | [96] |
| 2019 | ITO | | Poly-TPD | FPMAI-$MAPb_{0.6}Sn_{0.4}I_3$ | TPBi | LiF/Al | [105] |
| 2021 | ITO | | PEDOT:PSS | $CsSnI_3$ | B3PyMPM | LiF/Al | [108] |
| 2022 | ITO | PEDOT:PSS | VB-FPND | PEAI-modified $FAPbI_3$ | CN-T2T | LiF/Al | [103] |
| 2023 | ITO | | PEDOT:PSS /Poly-TPD/PVK | $CsPb(Cl_{1-x}Br_x)_3$: $Yb^{3+}$ NCs | TPBi | Liq/Al | [107] |

Among the various NIR emitting metal halides, $FAPbI_3$ has been mostly studied for NIR LEDs as the EML since it presents a narrow bandgap of ~1.48 eV and relatively high stability. With proper surface modification/passivation to improve carrier-transport properties and reduce nonradiative recombination, $FAPbI_3$ LEDs could be further optimized and reach an EQE of 23.8%. A noticeable example is the work of Huang et al. Here, the authors introduced amino-acid additives into $FAPbI_3$ precursor solutions, which could effectively passivate surface defects and suppress nonradiative recombination, thus obtaining $FAPbI_3$ NIR LEDs with a high EQE of 20.7% and a half-lifetime ($T_{50}$, the time taken for the device EQE to drop to 50% of its initial value) of 20 h .[5a] In another important work, Gao et al. found that defect passivation in $FAPbI_3$ through the introduction of organic molecules was dependent on the strength of hydrogen bonds in the latter.[5b] Therefore, they designed a passivation molecule (2,2′-[oxybis(ethylenoxy)]diethylamine, ODEA) with weak hydrogen bonds.[5b] The ODEA-treated NIR LED device displayed EQE up to 21.6%, with a EL peak at 800 nm and $T_{50}$ = 20 h, thus demonstrating the successful passivation strategy.[5b] Based on the positive results of Gao et al., other researchers started exploring specifically designed passivation molecules. For example, Cui et al. designed a multifunctional molecule (2-(4-(methylsulfonyl)phenyl)ethylamine, MSPE) which they added to the precursor of $FAPbI_3$ perovskites.[16d] The effects of MSPE were as follows: 1) Self-assembly of MSPE among discontinuous perovskite grains eliminated interfacial non-radiative recombination; 2) MSPE could act as a barrier for the quenching of luminescence at the interface between charge-transport layers and perovskite (**Figure 9**a). The MSPE-based device emitting NIR emission at 800 nm, exhibited a peak EQE of 23.8% at 33 mA $cm^{-2}$ and a $T_{50}$ of 32 h (Figure 9b, c).[16d]

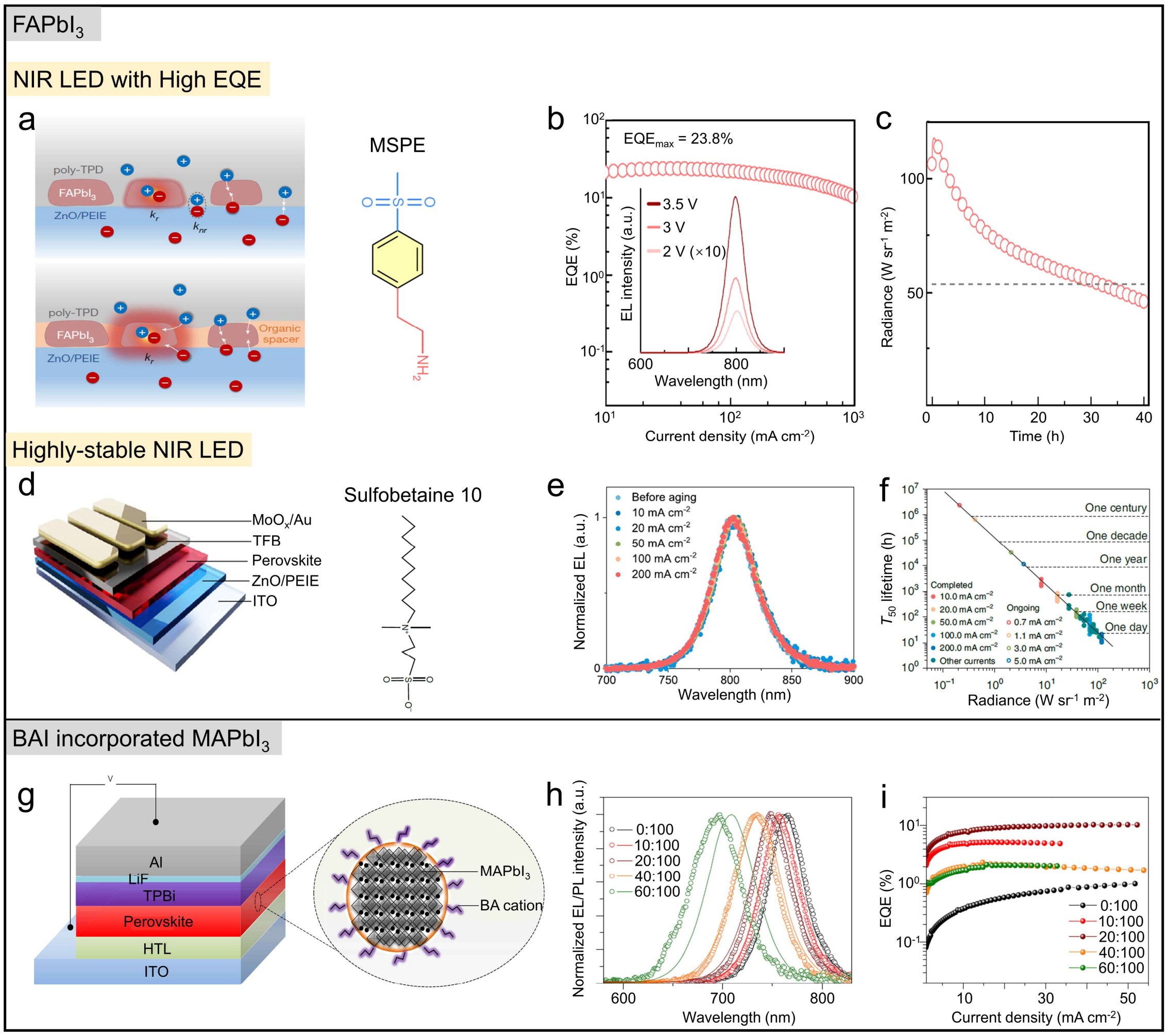

**Figure 9. NIR LED based on FAPbI₃. NIR LED with high EQE:** (a) Illustration of the LED device with control and MSPE-based perovskite as emitting layers and the molecular structure of MSPE. (b) EQE versus current density and EL spectra at different biases. (c) Operational stability of a MSPE-based LED. Reproduced (Adapted) with permission.[16d] Copyright 2023, Springer Nature. **Ultrastable NIR LED:** (d) Device structure and the molecular structure of SFB10. (e) EL spectra of SFB10-stabilized LEDs without and with accelerated aging tests at different current densities. (f) $T_{50}$ lifetimes curves as a function of initial radiance. Reproduced (Adapted) with permission.[94b] Copyright 2022, Springer Nature. **NIR LED based on $MAPbI_3$.** (g) Device structure and a sketch of $MAPbI_3$ with BA cations at the surface. (h) EL (circles), PL (lines) spectra and (i) EQE curves with different molar ratios of BAI:$MAPbI_3$. Reproduced (Adapted) with permission.[96] Copyright 2017, Springer Nature.

Despite the significant developments on the side of device efficiency, the poor stability of $FAPbI_3$ perovskite LEDs arising from the ionic nature of the crystals remains a critical challenge. The key issues limiting the device operation lifetime are the intrinsic instability of the emissive layers, defects in metal halide perovskite film and interface, ion migration under electric fields and the diffusion of metallic species. NIR emitting hybrid perovskites layers can easily decompose at high temperatures or in a high-humidity environment due to their strongly polar lattice and the volatility of the organic A cation. In fact, in a polycrystalline film during the crystallization process, point defects including vacancies, interstitial, anti-site and structural defects (including dislocations and grain boundaries) can form due to the small defect formation energy; In NC films, the ligands can detach from the surface of NCs during the film fabrication, generating defects. Those defects can act as non-radiative recombination centers and lead to a local increase of temperature, in turn promoting the decomposition of hybrid perovskites. Besides, the activation energy of ion migration is low, especially for halide ions, so it is easy to induce ion migration under the electric field, resulting in phase change and spectral instability. In addition, metallic species from the electrodes can diffuse into the emissive layers, quenching the luminescence and causing instability of EL intensity.[94b, 99, 101, 113] Many efforts have been made to resolve these issues. Ning et al. used 1,4-bis(aminomethyl)benzene molecules as bridging ligands to form the Dion-Jacobson (DJ) structure.[99] The NIR LED based on DJ structure showed an EQE of 5.2%, EL peak at 776 nm and maximum radiance of 88.5 W $sr^{-1}$ $m^{-2}$. The device possessed a large $T_{50}$ value of 100 hours.[99] Similarly, Gao et al. incorporated dicarboxylic acids into the $FAPbI_3$ precursor solution to eliminate defects and stabilize the perovskites layer,[101] thus leading to improved EQE (18.6%) and $T_{50}$ (682 h).[101] More recently, Di et al. used a dipolar molecular (sulfobetaine 10, SFB10) to stabilize $FAPbI_3$ emissive layers (Figure 9d) and thus increase the functional lifetimes ($T_{50}$) up to 11,539 h and 32,675 h at ~5.0 mA $cm^{-2}$ and ~3.2 mA $cm^{-2}$, respectively (Figure 9e, f, EQE = 22.8% and maximum radiance 278.9 W $sr^{-1}$ $m^{-2}$).[94b]

Although high efficiencies and stabilities have been achieved, the reported devices are of small area, generally several $mm^2$, and they cannot meet the needs of large area commercial use. Based on that, Tan et al. utilized a hole-transporting polymer, poly-TPD to build both small-area and large-area NIR devices. By employing poly-TPD, the authors optimized the charge balance in the NIR device.[16c] As a result, the small-area 2 × 2 $mm^2$ LED device exhibited an EQE of 20.2% with the EL peak at 799 nm and a radiance of 170 W $sr^{-1}$ $m^{-2}$, with a lifetime ($T_{80}$) of 20 h. Meanwhile, a 900 $mm^2$ large area NIR device was fabricated, exhibiting an EQE of 12.1%.[16c]

$MAPbI_3$ has also been explored in NIR LEDs, given its direct bandgap of 1.63 eV. However, the small exciton binding energy (16 meV) of $MAPbI_3$ leads to serious luminescence losses caused by the thermal ionization of the excitons. To address this issue, Choy et al.

introduced poly(2-ethyl-2-oxazoline) (PEtOz) into the $MAPbI_3$ to promote radiative recombination.[95] The NIR LEDs based on the optimized perovskite-PEtOz nanocomposite emission layer peaking at 760 nm achieved a radiance of 12.3 W $sr^{-1}$ $m^{-2}$ and an EQE of 0.76%.[95] Rand et al. added ammonium halides to the precursor solution, which could hinder the grain growth of $MAPbI_3$ crystallites and reduce film roughness (Figure 9g).[96] The NIR LED based on the N-butylammonium halides (BAI)-incorporated emissive layers that featured nanometre-sized grains and thus exhibited an improved EQE. The sample with molar ratio 20:100 of BAI:$MAPbI_3$ in the precursor reached the highest EQE: up to 10.4%, with an EL peak at 748 nm (Figure 9h, i).[96]

An efficient strategy to lower the bandgap of lead-based metal halides and reduce the lead toxicity is to replace lead (Pb) with tin (Sn). Tan et al. reported a $MASnI_3$-based NIR LED device with EL at 945 nm, EQE of 0.72% and a radiance of 3.4 W $sr^{-1}$ $m^{-2}$.[18] Wei et al. exploited the dendritic structure of $CsSnI_3$ films, reaching an EQE of 5.4% with EL peak at 932 nm and a $T_{50} = 23.6$ h.[108] Another example is the work of Rand et al., where the authors used Pb–Sn mixed halide perovskite as emitters combined with 4-fluorobenzylammonium iodide (FPMAI) to reduce the size of the crystal grains. The device with FPMAI-$MAPb_{0.6}Sn_{0.4}I_3$ emitting layer featured an EQE of 5% with an EL peak of 917 nm and a radiance of 2.7 W $Sr^{-1}$ $m^{-2}$ at 4.5 V.[105] However, when the molar ratio of Sn was less than 20%, mixed Pb–Sn perovskites became disordered with high trap density. To overcome this issue, Xiao et al. doped alkali cations to release microstrain and passivate the traps in mixed Pb–Sn perovskites.[111] The NIR device with a KI-doped $MAPb_{0.8}Sn_{0.2}I_3$ emissive layer showed an enhanced EQE of 9.6% with an EL peak at 868 nm.[111]

Perovskites with mixed A-site cations and halides composition were also proposed as the emitting layer to improve the stability of organic-inorganic perovskites. For example, Kovalenko et al. used $Cs_xFA_{1-x}Pb(Br_{1-y}I_y)_3$ NCs with a narrow-band emission peaking at 735 nm to fabricate a NIR LED.[48] The NC-LED exhibited an EQE of 5.9% and a short $T_{50}$ of 30 s at 4.5 V.[48] Nazeeruddin et al. explored the triple-cation mixed perovskite (FA/MA/Cs) used in NIR LED and obtained an EQE of 9.23%, an EL peak at 750 nm and a maximum radiance of 93.34 W $sr^{-1}$ $cm^{-2}$. The triple-cation perovskite-based NIR LED exhibited a relatively high operational stability with lifetime ($T_{100}$) of 300 min.[97]

Only few studies were reported on the application of doped perovskites emitting layers in NIR LED. For example, $Bi^{3+}$ doped $MAPbI_3$ showed a new NIR PL at ~1140 nm from Bi doping, apart from the intrinsic PL centered at 782 nm.[37a] Song et al. fabricated an electrically-driven device with $Bi^{3+}$ doped $MAPbI_3$ as the emissive layer,[37a] but no EL could be observed at room temperature. At 84 K, an EL peaking at 1100 nm with a FWHM of 226 nm was measured, which could be attributed to the suppressed non-radiative channels at cryogenic temperatures. Owing to the operation difficulty, it is was not possible to measure the EQE of the NIR LED at 84 K.[37a] It is worth of note that a host with a high mobility of electrically excited carriers can increase the radiative recombination of an emission center and result in a high EQE in LEDs. $CsPbCl_3$ could be used as a host with high mobility and a sensitizer for the NIR-emitting dopant $Yb^{3+}$ ions. Miyasaka et al. used a $Yb^{3+}$ doped $CsPbCl_3$ film as the emitting layer to fabricate a NIR LED.[106] The demonstrated NIR LED exhibited a EQE of 5.9% with an EL peak of 984 nm and a lifetime ($T_{50}$) of 58 h at 0.827 mA $cm^{-2}$.[106]

### 6.2 On-chip downshifting layers and imaging

NIR LEDs based on lead free metal halides as downshifting layers have also being tested.[24c, 30d, 32a, 39b, 63c, 65, 68, 70, 73a, 73c, 76c, 114] In some cases, NIR emitting metal halides have been mixed with other phosphors, thus obtaining a broadband white-NIR LED.[32a, 73a, 76c] For example, Wang et al. coated $Cs_2NaBi_{0.86}Er_{0.14}Cl_6$:$Mn^{2+}$ (red and NIR), $\beta$-SiAlON:$Eu^{2+}$ (green) and $BaMgAl_{10}O_{17}$:$Eu^{2+}$ (blue) phosphors on a commercial 365 nm UV chip, obtaining a white-NIR LED (**Figure 10**a, b).[73a] The constructed LED emitted both white light and NIR light at 1540 nm from $Er^{3+}$, extending the emission spectrum to the optical telecommunication range.[73a]

More often, NIR emitting metal halides are directly coated on the UV chip. For example, Liu et al. fabricated a UV to NIR downshifted LED by combining $Cs_2MoCl_6$ or $Cs_2WCl_6$ film with a commercial UV chip (Figure 10c, d).[30d] The obtained device can be used in blood vessels imaging owing to the deep tissue penetration of NIR emission. Blood vessels are in a layer of subcutaneous adipose tissue at a depth of at a depth of 1 mm to several millimeters. Under the skin surface above the subcutaneous adipose is epidermis (~0.1 mm) and dermis (~1 mm). The major chromophores in epidermis, dermis and subcutaneous adipose tissues include melanin, blood (mainly hemoglobin), lipids and water.[115] In the visible light region, the melanin and hemoglobin are both highly absorptive, decreasing the tissue penetration. In the red and NIR part of the spectrum, the tissues of the skin show relatively good transparency, while the hemoglobin in the blood is highly absorptive in the NIR region (700–1000 nm).[116] Thus, under NIR light illumination, a relatively high contrast between the blood vessels and surrounding tissues is present and blood vessels imaging can be achieved. By employing $Cs_2MoCl_6$ or $Cs_2WCl_6$ based UV to NIR downshifted LEDs as light source (NIR light at ~950 nm), images of the blood vessels near the skin surface were clearly captured by a NIR camera (Figure 10e).[30d] Chen et al. utilized a $Cs_2AgInCl_6$:Bi/Ln-in-glass composite and a commercial 350 nm UV chip to fabricate a Vis–NIR ultra-broadband LED.[63c] The EL spectra exhibited a narrow UV emission from the chip, a broadband visible emission from STE, and several narrow-band emissions from the *f–f* transitions of $Er^{3+}$, $Nd^{3+}$, $Yb^{3+}$ and $Tm^{3+}$ (Figure 10f). The LED device showed good stability and its PL intensity remained unchanged under exposure in air for 1200 h. The fabricated LED was also utilized for nondestructive absorption spectra analyses. As shown in Figure 10g, the fruit requiring testing was put in an integrating sphere equipped with the fabricated LED and a spectrometer.[63c] The absorption spectrum (ΔPL) of the substance was obtained by the following formula (Figure 10h):

$$\Delta PL = PL(LED) - PL(LED + substance)$$

Those examples indicate the promising applications of the NIR LEDs based on lead free NIR emitting metal halides in medical imaging and non-destructive testing.

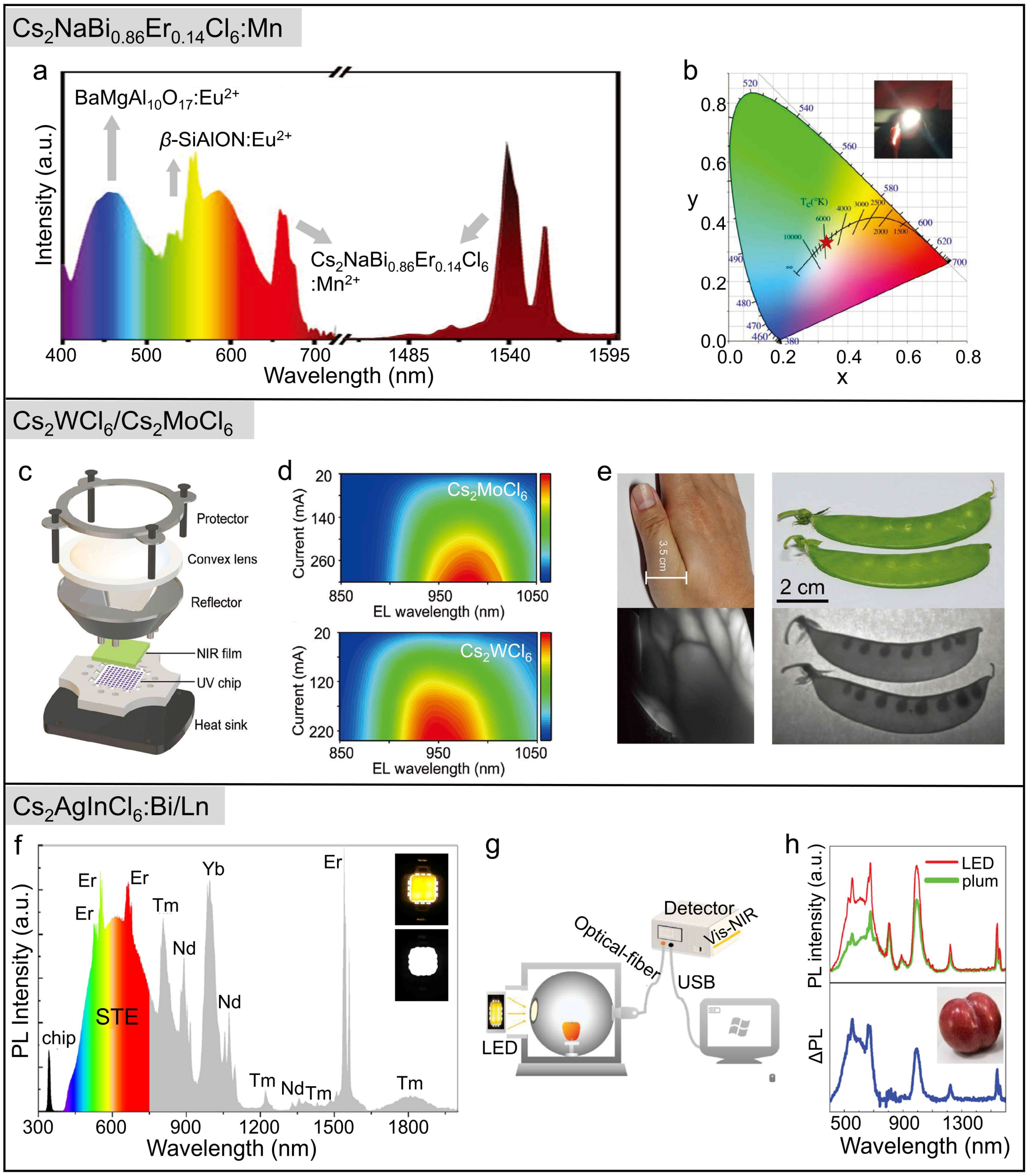

**Figure 10. WLED fabricated with $Cs_2NaBi_{0.86}Er_{0.14}Cl_6$:0.42%Mn as the red emitter:** (a) EL spectra. (b) CIE coordinates. Reproduced (Adapted) with permission.[73a] Copyright 2022, Wiley. **$Cs_2MoCl_6$-based or $Cs_2WCl_6$-based NIR emitting LEDs:** (c) Schematic of LED device. (d) Pseudocolor map of the PL spectra. (e) Imaging of blood vessels and plant seeds. Reproduced (Adapted) with permission.[30d] Copyright 2022, Wiley. **NIR emitting LED based on $Cs_2AgInCl_6$:Bi/Ln@glass composite.** (f) EL spectrum of the LED device. Inset shows photos of the constructed LED device recorded by the visible camera and NIR camera. (g) Schematic of the experimental setup to measure the absorption of fruits. (h) PL spectra with/without plum inside the integrated sphere using LED as the light source. Reproduced (Adapted) under the terms of the CC-BY Creative Commons Attribution 4.0 International License.[63c] Copyright 2022, The Authors, published by Springer Nature.

## 7. Conclusion and Outlook

In summary, we have provided an overview of different NIR emitting metal halides covering their structural and electronic features. Various strategies to achieve NIR emission have been discussed and their current applications in NIR LEDs, bio-imaging, non-destructive testing have been summarized. It is clear that significant progress has been made in the past few years and this class of materials is emerging as promising candidates for NIR light sources. Nonetheless, the development of NIR emitting metal halides with high efficiency, broadband emission and robust stability is still ongoing, especially in the NIR-II region. Since the quality of materials plays a critical role in the device performance, the realization of commercial applications for NIR emitting metal halides still has a long way to go.

For $ABX_3$-type lead-based metal halide perovskites with NIR emission, such as (MA, FA, Cs)$PbI_3$, their emissions are narrow-band and their spectra cover the visible region. For fluorescence imaging in biomedical applications, the emission in the NIR-II (1000–1700 nm) window is desirable compared to that in NIR-I region (750–950 nm). That is because the NIR-II emission can reduce tissue scattering and autofluorescence and thus achieve deep tissue penetration and high spatial resolution. To overcome the limited wavelength of emission (and also lead toxicity), $Sn^{2+}$ ions have been introduced as substitutes for $Pb^{2+}$ ions. This has been shown to narrow the bandgap and extend the PL to NIR-II region with no visible background emission. However, tin-based metal halide perovskites with NIR emission are limited by poor stability owing to ease of oxidization of $Sn^{2+}$ and the corresponding NIR LEDs suffer the same problem. Yet, simply inhibiting the oxidation of $Sn^{2+}$ could not lead to considerable improvements in the performance of tin-based perovskite NIR LED, since they additionally suffer from high hole concentration and mobility, and a balanced injections of electrons and holes is very important to ensure effective radiative recombination. Hence, for tin-based NIR LEDs, both suppressing the oxidation of $Sn^{2+}$ and increasing electron injection are highly desirable.

$Ln^{3+}$ ion doping seems to be an effective method to extend the emission into the NIR-II region. Yet, it is unsatisfactory that $Ln^{3+}$ induced NIR emissions are characterized by sharp spectral features, narrow-band light absorption and relatively low PLQY. Furthermore, although for some $Ln^{3+}$ doped metal halides, such as $Ln^{3+}$ doped double perovskite, the NIR emission from $Ln^{3+}$ ions could be sensitized by the host, the excitation is still limited to high energy (< 350 nm, it cannot match commercial blue LED chip, $\lambda = 450$ nm) and can be extended to the red-spectral region only in a few cases. Besides, apart from the NIR PL with high PLQY owing to the quantum cutting effect in $Yb^{3+}$ doped $CsPbCl_3$, it is still challenging to achieve efficient NIR emission from $Ln^{3+}$ ions in metal halides. That is due to a lack of suitable luminescent centers that can effectively absorb the excitation energy and transfer energy to $Ln^{3+}$ ions in those metal halides with indirect or direct bandgaps with parity-forbidden transition. Therefore, it is imperative to identify more effective methods to trigger the sensitization of $Ln^{3+}$ ions in metal halides and obtain more efficient NIR luminescence. Searching for proper co-dopants acting as an efficient energy transfer bridge could be a possible direction to obtain efficient NIR luminescence.

Broadband NIR emission could be achieved by $Cr^{3+}$, $Sb^{3+}$ and $Bi^{3+}$ ions doping. As discussed in this review, $Cr^{3+}$ ions have been doped into $Cs_2(Ag,Na)InCl_6$, delivering a NIR emission with a large FWHM of 193 nm. However, the amount of $Cr^{3+}$ that could be introduced in $Cs_2(Ag,Na)InCl_6$ NCs was low (< 2%), and this is most likely the reason for the low NIR PLQY (~20% both for bulk materials and NCs). $Sb^{3+}$ and $Bi^{3+}$ are both n$s^2$ ions that are widely used as activators to obtain broadband emission from STE. The emission of n$s^2$ ions doped metal halides can be tuned from blue to NIR in different hosts. But how can we choose the appropriate structure to get NIR emission from n$s^2$ ions doping? In most cases, the emission band from n$s^2$ ions doping covers mainly the visible region rather than the NIR one, and this is harmful for medical imaging with high signal-to-background ratio due to the photon scattering and tissue autofluorescence interference. Therefore, one needs to explore strategies to increase the doping amount and develop a deeper understanding for the localized environment of n$s^2$ ions to achieve NIR luminescence.

Thermal, phase and luminescence stability of NIR emitting metal halides should be improved if one wants to fabricate high-performance NIR LEDs. NIR emitting lead-based perovskites generally suffer from thermal degradation at high temperatures. Ion doping/alloying (A-site or halide ions) and molecular additives (dicarboxylic acids, sulfobetaine 10, 2-(4-(methylsulfonyl)phenyl)ethylamine (MSPE)) can improve their thermal stability.[16d, 48, 94b, 101] As for NIR emitting lead-free metal halides, all-inorganic compounds have better phase stability than the corresponding organic-inorganic counterparts. One exception is $(TMEDA)(Sb,Bi)I_5$ SCs, due to the presence of hydrogen bonds between the organic cations and the inorganic component (chains of octahedra), leading to a relatively stable structural framework.[30b] Among the reported all-inorganic NIR emitting lead-free metal halides, some display a good thermal performance. For example, $Cs_2NaBi_{0.86}Er_{0.14}Cl_6$:0.42%Mn MCs, when heated at 150 ºC, could retain ~90% PL intensity of the initial value after 550 h.[73a] The PL of $Cs_2ZnCl_4$:$Sb^{3+}$ SCs at 150 ºC preserved 73% of its RT intensity value.[39a] The PL intensity from $Bi^{3+}$–$Er^{3+}$ co-doped $Cs_2AgInCl_6$ remained almost unchanged in the temperature range from 5.7 K to 300 K.[24a] There are also reports of anti-thermal quenching, as for $Cs_2NaEr_{0.4}Yb_{0.6}Cl_6$ SCs (emission at 995 nm, and intensity increasing from 80 K to 550 K),[73c] and zero-thermal quenching: $Cs_4Cd_{1-x}Mn_xSb_2Cl_{12}$:$Yb^{3+}$ ($x$ = 0.3, 0.6) (from 80 K to 500 K).[76c] Despite these encouraging cases, most NIR emitting metal halides are vulnerable to serious thermal-quenching and cannot satisfy the requirements of commercial application.[32b, 63c, 68] The temperature of phosphors in LEDs under a prolonged operation can be up to 100–200 ºC, due to the heat dissipation from the underlying chip.[117] Hence, improving the thermal stability of NIR emitting metal halides is a critical issue. (1) Ion doping/co-doping is an alternative strategy that can lead to weakening of electron–phonon coupling or to an increased energy transfer process. In fact, in the case of STE induced NIR emission, the incorporation of some specific ions can enhance the exciton binding energy, reduce the electron–phonon interaction and contribute to reduce thermal quenching. For example, in NIR emitting $Cs_2Ag_{0.05}Na_{0.95}BiCl_6$ microcrystals, the incorporated $Na^+$ separates $[AgCl_6]^{5-}$ octahedra by means of $[NaCl_6]^{5-}$ octahedra acting as barriers to confine the spatial distribution of the STE state, mainly by promoting the localization of holes, and weakening the strong electron-phonon coupling of the corresponding unalloyed system.[30f] For the cases of $Ln^{3+}$ ion induced NIR emission, ions doping/co-doping can further promote energy transfer. Based on the energy difference between the excited states of the interacting ions, energy transfer should be strongly dependent on the temperature, with higher temperatures improving its efficiency. For example, in $Yb^{3+}$ doped $Cs_4Cd_{1-x}Mn_xSb_2Cl_{12}$ ($x$ = 0.3) microcrystals, NIR emission from $Yb^{3+}$ ions displayed zero-thermal quenching from 80 to 500 K and higher $Mn^{2+}$ concentration led PL quenching temperatures higher than 500 K due to a more efficient energy transfer from $Mn^{2+}$ to $Yb^{3+}$ ions.[76c] (2) Another possible venue is to learn from the design of phosphors in glass (PiG), incorporation of metal halides in glass can be an effective method to improve their thermal stability.[117] A proper glass with low-melting temperature should be considered to avoid high-temperature co-sintering. For example, a thermally-stable $Ln^{3+}$ doped double perovskites embedded in glass was designed by a low temperature co-sintering method and no obvious changes of the $Ln^{3+}$ emissions occurred after continuous operation for 48 h.[63c]

From the perspective of electrically-driven NIR LEDs, most emissive layers are lead-based perovskites. To address the toxicity issue of $Pb^{2+}$, the development of lead-free electrically-driven NIR LEDs becomes an inevitable trend. Among them, tin-based electrically-driven LEDs stand out due to the low bandgap, high hole mobilities and the similar characteristics of $Sn^{2+}$ with $Pb^{2+}$. Nevertheless, their performances are not comparable to those of lead-based NIR LEDs, mainly from the facile $Sn^{2+}$ oxidation. In fact, $Sn^{2+}$ ions can be easily oxidized to $Sn^{4+}$ ions, resulting in the formation of Sn vacancies, high hole concentration and unbalanced charge injection. Those factors thus lead to undesired non-radiative recombination and poor device performance. Although many other lead-free NIR emitting metal halides with excellent optical properties have been reported, studies on electrically-driven LEDs are scarce.[118] Most reported NIR emitting lead-free metal halides are characterized by large bandgaps and low-dimensional structures, limiting their carrier mobility. Also, low solubilities of some precursors in common organic solvents result in poor film morphology for bulk perovskites. On the other hand, although the bulk counterparts show high PLQY, their nanocrystals present low PLQYs due to the lack of defect tolerance. Overall, the development of lead-free electrically-driven NIR LEDs still has a long way to go, as significant issues need to be overcome in this direction. More recently, Gao et al. developed bright and stable lead-free NIR LEDs based on $CsSnI_3$ emitting at 948 nm with a radiance of 226 W $Sr^{-1}$ $m^{-2}$ and $T_{50}$ of 39.5 h at a constant current density of 100 mA $cm^{-2}$. The improved performance was achieved by the manipulation of intrinsic p-doping density and trap density through controlling the crystallization process under tin-rich condition, further promoting the development of lead-free electrically-driven NIR LEDs.[110] We hope that the challenges and opportunities will stimulate more studies in this emerging field and facilitate the applications of NIR-emitting metal halides.

**Acknowledgements**

The work was supported by the National Key R&D Program of China (No. 2022YFB2803900), the National Natural Science Foundation of China (No. 12074347, 12304458, 12304473 and 52302171), Support Program for Scientific and Technological Innovation Teams of Higher Education in Henan Province (No. 231RTSTHN012) and the Key Project for Science and Technology Development of Henan Province (No. 232102231039). L. M. acknowledges funding from the programme MiSE-ENEA under the Grant "Italian Energy Materials Acceleration Platform – IEMAP" and from the European Research Council though the ERC Advaced Grant NEHA (contract n. 101095974).